\documentclass{article} 
\usepackage{iclr2027_conference,times}

\usepackage{amsmath,amsfonts,bm}

\def\eqref#1{equation~\ref{#1}}

\def\1{\bm{1}}

\DeclareMathAlphabet{\mathsfit}{\encodingdefault}{\sfdefault}{m}{sl}
\SetMathAlphabet{\mathsfit}{bold}{\encodingdefault}{\sfdefault}{bx}{n}

\usepackage{kotex}
\usepackage{hyperref}
\usepackage{url}
\usepackage{bbm}
\usepackage{algorithmic,algorithm}
\usepackage{subcaption}
\usepackage{booktabs}
\usepackage{caption}
\usepackage{makecell, multirow}
\usepackage{tikz}
\usetikzlibrary{positioning,fit,calc,arrows.meta,decorations.pathreplacing}

\usepackage{graphicx}               
\usepackage{tabularx}               
\usepackage{booktabs}
\usepackage{amsmath}
\usepackage{stmaryrd}
\usepackage{xcolor}
\usepackage{soul}

\newcolumntype{C}{>{\centering\arraybackslash}X}
\usepackage{multirow}               
\usepackage{diagbox}                
\usepackage{hhline}                 
\usepackage{color}                  
\usepackage{amsmath}                
\usepackage{amssymb}                
\usepackage{mathtools}              
\usepackage{enumitem}               

\usepackage{booktabs}
\usepackage{extarrows}
\usepackage{makecell}

\usepackage{soul}

\definecolor{RoseQuartzBg}{HTML}{F7CAC9}
\definecolor{RoseQuartz}{HTML}{F5A798}
\definecolor{Serenity}{HTML}{92A8D1}
\definecolor{OrangeRed}{rgb}{1.0, 0.27, 0.0}
\definecolor{Red}{rgb}{1.0, 0.0, 0.0}
\definecolor{Turquoise}{HTML}{0F4C81}
\usepackage{xparse}

\NewDocumentCommand{\jy}{ mO{} }{\textcolor{RoseQuartz}{\textsuperscript{\textit{jy}}\textsf{\textbf{\small[#1]}}}}
\NewDocumentCommand{\mh}{ mO{} }{\textcolor{blue}{\textsuperscript{\textit{mh}}\textsf{\textbf{\small[#1]}}}}

\newcommand{\ParaSet}{\texttt{ParaSet}}

\title{Scoring a Set, Not Summing Passage Scores: Effective and Efficient Set Retrieval}

\author{Mooho Song, Jay-Yoon Lee\thanks{Corresponding Author}\\
Graduate School of Data Science\\
Seoul National University\\
\texttt{\{anmh9161, lee.jayyoon\}@snu.ac.kr}
}

\iclrfinalcopy 
\begin{document}

\maketitle
\lhead{Under review at ICLR 2027} 

\begin{abstract}
Multi-hop question answering requires retrieving multiple evidence passages whose usefulness often depends on one another.
Conventional retrievers either rank passages independently or construct evidence sequentially through locally supervised next-passage decisions.
Sequential conditioning captures some cross-passage dependencies, but its local extension scores do not provide a common criterion for comparing complete evidence sets of different compositions and sizes.
Existing multi-hop retrievers therefore avoid directly learning a query--set compatibility function over complete evidence sets, as the exponentially large set space is daunting to cover during learning and impractical to search at inference time.
To overcome these challenges, we formulate multi-hop retrieval by directly ranking candidate evidence sets with an energy-based query--set compatibility score $s_\theta(q,S)$, learned from informative contrasts without exhaustive coverage or normalization of the combinatorial set space.
Given a gold evidence set, we automatically generate contrasts by adding, removing, or replacing passages, yielding rich set-level supervision without additional annotation.
To make search over this combinatorial space practical at inference time, we introduce a novel retrieve-and-rerank framework over the set space: \ParaSet{}, a lightweight scorer over precomputed passage representations for efficient set exploration, and \texttt{SetCE}, an expressive cross-encoder for set reranking.
Across three multi-hop QA benchmarks and two first-stage retrievers, set-level retrieval consistently provides a complementary signal to passage-level relevance, becoming relatively more effective when more hops are required to reach evidence from the query.
Motivated by this complementarity, combining the two signals further improves downstream QA performance, outperforming both deeper passage-level retrieval and an ensemble of distinct passage-level retrievers.
Finally, without invoking an LLM during retrieval, our method remains competitive with LLM-assisted evidence-selection pipelines while achieving up to $15\times$ faster evidence selection.
\end{abstract}

\section{Introduction}
\label{section:introduction}

Document retrieval is commonly formulated as ranking individual passages by their respective relevance to a query.
This formulation is effective when useful evidence can be identified
independently, but many reasoning-intensive tasks, such as multi-hop
question answering, require multiple passages whose usefulness depends on
one another.
For example, a supporting passage may have only weak direct relevance to
the original query, yet become essential when combined with another
passage that establishes an intermediate entity or relation.
Such cases suggest that multi-hop retrieval involves not only
\emph{query--passage relevance}, but also \emph{query--set compatibility}:
whether a collection of passages jointly forms useful evidence for
answering the query.

Existing retrieval paradigms capture such interactions in different ways.
Passage-level retrievers learn scores over individual query--passage pairs,
while sequential retrievers condition each next-passage decision on
previously selected evidence~\citep{e2e,multihop_dense_retrieval}.
Sequential conditioning captures some cross-passage dependencies, but still
evaluates one extension at a time.
A passage that appears useful under the current prefix may later become
redundant or incompatible once additional evidence is considered, while a
passage with weak local relevance may become useful only as part of a
larger evidence set.
Moreover, sequential retrieval typically extends a hypothesis by adding passages, so an earlier passage cannot be retrospectively removed if later evidence reveals it to be unhelpful.
Recent set-selection methods can explicitly reason over multiple passages,
but rely heavily on LLM decoding~\citep{setr,arm,optiset}.
What these paradigms lack is an explicit learned
evaluator that assigns a directly comparable scalar score to an arbitrary
candidate evidence set.

Instead of only learning $s(q,d)$ or locally supervised extension scores,
we learn a global compatibility function $s_\theta(q,S)$ that directly
compares candidate sets with different compositions and cardinalities.
The key distinction in our formulation is that the complete
evidence set itself serves as the learned scoring unit, rather than merely
as the output of a multi-passage selection procedure.


Building on the set-level energy formulation introduced for consistency
verification~\citep{set-consistency}, we extend this perspective to
multi-hop retrieval by learning a query--set compatibility score that
directly ranks complete and compatible evidence sets above incomplete,
noisy, or corrupted alternatives.
Direct set-level learning, however, introduces a combinatorial challenge
at both training and inference time.
A candidate pool of $k_b$ passages induces $2^{k_b}-1$ non-empty sets,
making it impractical to exhaustively enumerate the output space for
supervision or normalization.
Rather than attempting to cover this space explicitly, we learn relative
preferences from informative contrasts around the gold evidence set.
Specifically, given a gold set, we automatically construct alternatives by
adding distractors, removing required evidence, or replacing required
evidence with distractors.
These structured perturbations provide rich set-level supervision without
requiring exhaustive enumeration or additional annotation.
The same combinatorial space also makes exhaustive inference infeasible.
We therefore factor inference into lightweight candidate-set exploration
and expressive reranking.
\ParaSet{} (Parallel Set Scorer) is a shallow late-interaction scorer over
precomputed bi-encoder representations that efficiently evaluates sets
encountered during approximate beam-search exploration.
The top-ranked proposals are then reranked by \texttt{SetCE}
(Set-level Cross Encoder), a more expressive cross-encoder trained with
the same set-level compatibility objective.

Experiments across three multi-hop QA benchmarks show that this set-level
formulation consistently benefits retrieval, but also reveal that it is
not simply a stronger replacement for passage-level relevance.
Its relative advantage tends to increase when a more hops are required to reach evidence from the query, suggesting that
passage-level relevance and set-level compatibility capture complementary evidence structures.
Our analysis further indicates that recovering the complete evidence set is
particularly important in multi-hop QA (Appendix \ref{appendix:retrieval_metric_correlation}). 
This structural complementarity motivates combining passage-level and set-level retrieval.
The combined system consistently improves downstream QA over deeper
passage-level retrieval and also outperforms combining two distinct
passage-level scorers.
Moreover, despite using no LLM during retrieval, its end-to-end performance remains competitive with agentic llm pipline while reducing retrieval time up to $15\times$.

Our contributions are summarized as follows:

\begin{itemize}[leftmargin=*]

    \item \textbf{Direct query--set scorer for multi-hop retrieval.}
    We propose set-level compatibility learning to train an energy-based model
    that measures compatibility between a query and a complete evidence set,
    enabling direct ranking of evidence sets rather than relying on arbitrary
    top-$k$ selection of independently scored passages or locally supervised
    sequential decisions, where earlier selections persist within a retained
    hypothesis even if they become unhelpful after additional evidence is
    observed.
    \item \textbf{Set-level retriever and reranker.}
    In order to efficiently search the combinatorial space, we propose a set-level retrieve-and-rerank pipeline with two set-level scorers: \ParaSet{} for efficient candidate exploration and \texttt{SetCE} for expressive set reranking. Analogous to passage-level bi- and cross-encoders, \ParaSet{} retrieves set-level candidates which \texttt{SetCE} then reranks. Experiments show that the proposed framework retrieves better evidence sets than using \texttt{SetCE} alone, while also being consistently faster.

    \item \textbf{Structural complementarity with passage-level retrieval.}
    By defining the \emph{Evidence Expansion Ratio} (EER), we show that the
    relative benefit of set-level retrieval tends to increase as a smaller
    fraction of the required evidence is directly connected to the query.
    Motivated by this complementarity, combining set-level compatibility
    with passage-level relevance further improves downstream QA over both
    deeper passage retrieval and an ensemble of two passage-level scorers.

\end{itemize}

\section{Related Work}
\label{section:related work}

\begin{figure*}[t]
    \centering
    \includegraphics[width=0.95\textwidth]{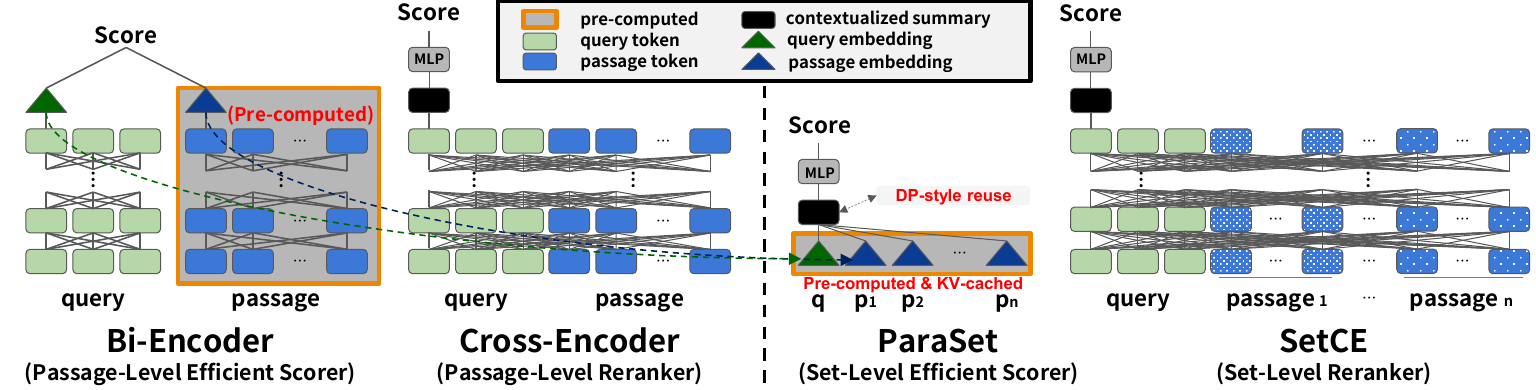}
    \caption{
    Overview of passage-level and set-level scoring.
    Passage-level retrievers score individual query--passage pairs,
    whereas our models directly score the compatibility between a query
    and a candidate passage set.
    \ParaSet{} reuses precomputed bi-encoder representations and applies
     set interaction for efficient candidate exploration,
    while \texttt{SetCE} jointly encodes the query and passages for
    expressive set-level reranking.
    }
    \label{fig:paraset_setce_overview}
\end{figure*}

\subsection{Multi-Passage Retrieval for Multi-hop QA}

A common approach to multi-hop QA is to retrieve multiple passages by independently ranking their relevance to a query.
Dense retrievers and cross-encoders typically return the top-$k$ passages under individual query--passage scores~\citep{dpr,fusion-in-decoder,rag}.
While simple and effective, this formulation does not explicitly account for dependencies among retrieved passages.
Consequently, evidence that is relevant through its connection to other evidence can be difficult to identify without direct retrieval cues in the query.

Another line of work performs retrieval iteratively, conditioning each
step on previously retrieved evidence or an evolving reasoning state
~\citep{multihop_dense_retrieval,react,self-ask,ircot,iterdrag}.
For example, \citet{e2e} use a cross-encoder with beam search to construct
multi-hop retrieval chains using next-passage supervision.
While such approaches model cross-passage dependencies through retrieval history, their learned scores are supervised over local retrieval decisions.
As each evidence set is usually scored by  local decision scores, earlier selections are not easily reevaluated in light of subsequently retrieved evidence.

In contrast, our scorer is directly supervised to evaluate a complete evidence set as a whole. Our \textit{set-level beam search} expands sets incrementally but scores each expanded set as a whole, rather than accumulating local decision scores.

\subsection{Set Selection and Structured Retrieval}

Recent work has moved beyond independent query--passage matching and
explicitly considers interactions among multiple passages.
SetR~\citep{setr}, OptiSet~\citep{optiset}, and ARM~\citep{arm} all perform
structured evidence selection through generation- or decoding-based
procedures, using expensive LLM reasoning, generative set selection, or constrained
decoding with additional optimization.
These approaches can capture dependencies among passages, but their set
preferences are tied to the corresponding selection procedure rather than
exposed as a standalone scalar function that can directly score arbitrary
query--set pairs.

Our work instead learns an explicit compatibility function
$s_\theta(q,S)$ that assigns a scalar score to any candidate evidence set.
The scorer is trained from gold evidence and systematic set perturbations,
and can be reused directly for candidate comparison, search, and reranking.
Unlike generation-based set selection, it also avoids LLM inference during
evidence selection, substantially reducing retrieval-time computation while
remaining competitive in downstream QA
(Table~\ref{tab:llm_assisted_comparison}).

\section{Set-Level Query--Set Scoring}
\label{section:set-cmc-retrieval}

Given a query $q$, let
$
\mathcal{D}_q
=
\{d_1,\ldots,d_{k_b}\}
$
denote the candidate passage pool produced by a first-stage retriever.
Conventional passage-level retrieval assigns a relevance score to each
$d_i$ independently.
Instead, we aim to learn
$
s_\theta(q,S),
\
\emptyset \neq S \subseteq \mathcal{D}_q,
$
which directly evaluates the usefulness of a candidate passage set $S$
for answering $q$.

This formulation raises two distinct challenges.
First, the scorer must learn comparable preferences over sets with
different compositions and cardinalities.
The appropriate amount of evidence can vary across queries, so set quality
should not be tied to a fixed number of retrieved passages.
Second, inference requires searching over as many as
$2^{k_b}-1$ non-empty subsets, making exhaustive evaluation impractical.
We address these challenges in turn.
To address the first, Section~\ref{subsection:set-level-compatibility-learning}
introduces a model-agnostic objective that learns comparable
query--set compatibility scores across candidate sets with different
compositions and cardinalities.
Section~\ref{subsection:set-level-scorers} then instantiates this objective
with two scorers at different points of the computation--expressivity
trade-off:
\ParaSet{}, a lightweight scorer for efficient candidate-set exploration,
and \texttt{SetCE}, an expressive cross-encoder for fine-grained set scoring.
To address the second challenge,
Section~\ref{subsection:set-level-inference} introduces
\textit{set-level beam search}, an approximate inference procedure for
searching the combinatorial set space, together with our
\textit{set-level retrieve-and-rerank pipeline}.


\subsection{Set-Level Compatibility Learning}
\label{subsection:set-level-compatibility-learning}

Let the gold evidence set for query $q$ be
$
\mathcal{G}_q=\{d_1^+,\ldots,d_m^+\}.
$
Set retrieval induces a combinatorial output space containing sets with
different compositions and cardinalities.
Following an energy-based structured prediction perspective
~\citep{set-consistency}, we therefore learn an unnormalized compatibility
score $s_\theta(q,S)$ for arbitrary query--set pairs, allowing candidate
sets to be compared through relative preferences without normalizing over
the full set space.
Rather than treating set retrieval as a binary distinction between the
gold set and arbitrary negatives, we learn a graded preference over
candidate evidence sets.
This allows the scorer to distinguish not only correct from incorrect
sets, but also complete evidence from partially correct or unnecessarily
noisy alternatives.

\paragraph{Set-quality ordering.}
For multi-hop QA, evidence coverage is particularly important:
missing a required passage may prevent the downstream model from
completing the reasoning chain, whereas a limited amount of additional
context can often be tolerated.
We therefore rank candidate sets lexicographically by evidence recall
and precision with respect to the gold set $\mathcal{G}_q$, prioritizing
recall and using precision to break ties.
In other words, a set covering more required evidence is always preferred,
while among sets with the same evidence coverage, the one containing fewer
irrelevant passages is preferred.
For example, when
$\mathcal{G}_q=\{d_1^+,d_2^+\}$,
the ordering gives
$
\{d_1^+,d_2^+\}
\succ
\{d_1^+,d_2^+,d^-\}
\succ
\{d_1^+\}
\succ
\{d_1^+,d^-\}.
$
Thus, complete evidence is preferred over incomplete evidence even when
the former contains an additional distractor.

\paragraph{Training candidates.}
We construct candidate sets spanning different quality levels by
perturbing the gold evidence set.
Specifically, we generate sets by adding distractors, removing required
evidence, or replacing required evidence with distractors.
We additionally include in-batch negatives formed from evidence
associated with other queries.
These perturbations expose the model to differences in both evidence
coverage and noise, rather than training only against unrelated negative
passages.
Details are provided in
Appendix~\ref{appendix:set_candidate_construction}.

\paragraph{Compatibility objective.}
Let $\mathcal{C}_q$ denote the resulting candidate-set pool.
For two candidates $H$ and $L$ such that $H\succ L$, we define
$
\ell(H,L)
=
\left[
\gamma
-
s_\theta(q,H)
+
s_\theta(q,L)
\right]_+,
$
where $\gamma>0$ is a fixed ranking margin.
We impose two forms of supervision.
First, the complete gold set should outrank every alternative candidate.
Second, we encourage finer separation between neighboring quality levels.
The resulting objective is
\[
\mathcal{L}
=
\sum_q
\left[
\sum_{L\in\mathcal{C}_q\setminus\{\mathcal{G}_q\}}
\ell(\mathcal{G}_q,L)
+
\lambda
\sum_{(H,L)\in\mathcal{P}_{\mathrm{fine}}^q}
\ell(H,L)
\right],
\]
where $\mathcal{P}_{\mathrm{fine}}^q$ contains ordered pairs from
neighboring quality levels and $\lambda$ controls the strength of the
fine-grained ranking term.
Importantly, this objective makes no assumption about the architecture of
$s_\theta$.
We apply the same set-level supervision to both \ParaSet{}(\ref{subsection:paraset}) and
\texttt{SetCE}(\ref{subsection:setce}), allowing us to separate the notion of
\emph{what constitutes a good evidence set} from the architecture used to
score it.

\subsection{Set-Level Scorers}
\label{subsection:set-level-scorers}

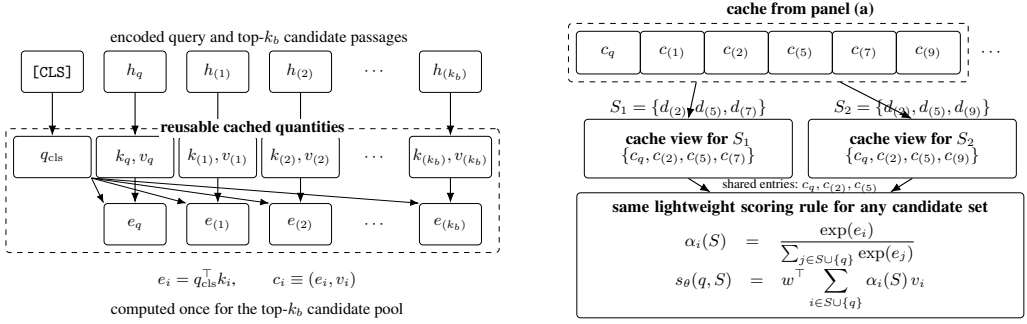
\begin{figure*}[t]
\centering

\begin{subfigure}[t]{0.47\textwidth}
\centering

\resizebox{\linewidth}{!}{%
\begin{tikzpicture}[
    >=Latex,
    font=\small,
    enc/.style={
        draw,
        rounded corners=2pt,
        align=center,
        minimum width=12mm,
        minimum height=8mm,
        inner sep=2pt
    },
    proj/.style={
        draw,
        rounded corners=2pt,
        align=center,
        minimum width=15mm,
        minimum height=8mm,
        inner sep=2pt
    },
    term/.style={
        draw,
        rounded corners=2pt,
        align=center,
        minimum width=12mm,
        minimum height=8mm,
        inner sep=2pt
    },
    group/.style={
        draw,
        dashed,
        rounded corners=2pt,
        inner sep=4pt
    },
    arrow/.style={->, semithick}
]

\path[use as bounding box]
(-0.5,0.8) rectangle (9.1,-5.25);

\node[font=\footnotesize] at (4.3,0.55)
{encoded query and top-$k_b$ candidate passages};

\node[enc] (cls) at (0.3,-0.1) {\texttt{[CLS]}};
\node[enc] (hq)  at (1.9,-0.1) {$h_q$};
\node[enc] (h1)  at (3.5,-0.1) {$h_{(1)}$};
\node[enc] (h2)  at (5.1,-0.1) {$h_{(2)}$};
\node          (hdots) at (6.55,-0.1) {$\cdots$};
\node[enc] (hK)  at (8.0,-0.1) {$h_{(k_b)}$};

\node[proj] (qcls) at (0.3,-1.75) {$q_{\mathrm{cls}}$};
\node[proj] (kvq)  at (1.9,-1.75) {$k_q,v_q$};
\node[proj] (kv1)  at (3.5,-1.75) {$k_{(1)},v_{(1)}$};
\node[proj] (kv2)  at (5.1,-1.75) {$k_{(2)},v_{(2)}$};
\node             (kvdots) at (6.55,-1.75) {$\cdots$};
\node[proj] (kvK)  at (8.0,-1.75) {$k_{(k_b)},v_{(k_b)}$};

\node[term] (eq) at (1.9,-3.05) {$e_q$};
\node[term] (e1) at (3.5,-3.05) {$e_{(1)}$};
\node[term] (e2) at (5.1,-3.05) {$e_{(2)}$};
\node           (edots) at (6.55,-3.05) {$\cdots$};
\node[term] (eK) at (8.0,-3.05) {$e_{(k_b)}$};

\draw[arrow] (cls) -- (qcls);
\draw[arrow] (hq)  -- (kvq);
\draw[arrow] (h1)  -- (kv1);
\draw[arrow] (h2)  -- (kv2);
\draw[arrow] (hK)  -- (kvK);

\draw[arrow] (kvq) -- (eq);
\draw[arrow] (kv1) -- (e1);
\draw[arrow] (kv2) -- (e2);
\draw[arrow] (kvK) -- (eK);

\draw[->, thin] (qcls.south east) -- (eq.north west);
\draw[->, thin] (qcls.south east) -- (e1.north west);
\draw[->, thin] (qcls.south east) -- (e2.north west);
\draw[->, thin] (qcls.south east) -- (eK.north west);

\node[
    group,
    fit=(qcls)(kvq)(kv1)(kv2)(kvdots)(kvK)
        (eq)(e1)(e2)(edots)(eK)
] (cachegroup) {};

\node[
    fill=white,
    inner sep=1pt,
    font=\footnotesize\bfseries
] at (cachegroup.north)
{reusable cached quantities};

\node[
    font=\footnotesize,
    align=center
] at (4.3,-4.10)
{
$e_i = q_{\mathrm{cls}}^\top k_i$,
\qquad
$c_i \equiv (e_i,v_i)$
};

\node[
    font=\footnotesize,
    align=center
] at (4.3,-4.72)
{
computed once for the top-$k_b$ candidate pool
};

\end{tikzpicture}%
}

\caption{
\textbf{What is cached?}
Reusable per-element quantities are computed once for the query and
top-$k_b$ candidate passages.
}
\label{fig:paraset_cache_a}

\end{subfigure}
\hfill
\begin{subfigure}[t]{0.50\textwidth}
\centering

\resizebox{\linewidth}{!}{%
\begin{tikzpicture}[
    >=Latex,
    font=\small,
    cache/.style={
        draw,
        rounded corners=2pt,
        align=center,
        minimum width=12mm,
        minimum height=8mm,
        inner sep=2pt
    },
    view/.style={
        draw,
        rounded corners=2pt,
        align=center,
        minimum width=39mm,
        minimum height=11mm,
        inner sep=3pt
    },
    score/.style={
        draw,
        rounded corners=2pt,
        align=center,
        text width=8.1cm,
        minimum height=15mm,
        inner sep=4pt
    },
    group/.style={
        draw,
        dashed,
        rounded corners=2pt,
        inner sep=4pt
    },
    arrow/.style={->, semithick}
]

\path[use as bounding box] (-0.4,0.8) rectangle (9.5,-5.55);

\node[font=\footnotesize\bfseries] at (4.5,0.6)
{cache from panel (a)};

\node[cache] (cq) at (0.9,-0.2) {$c_q$};
\node[cache] (c1) at (2.1,-0.2) {$c_{(1)}$};
\node[cache] (c2) at (3.3,-0.2) {$c_{(2)}$};
\node[cache] (c5) at (4.5,-0.2) {$c_{(5)}$};
\node[cache] (c7) at (5.7,-0.2) {$c_{(7)}$};
\node[cache] (c9) at (6.9,-0.2) {$c_{(9)}$};
\node at (8.1,-0.2) {$\cdots$};

\node[group, fit=(cq)(c1)(c2)(c5)(c7)(c9)] (bank) {};


\node[font=\footnotesize] at (2.4,-1.25)
{$S_1=\{d_{(2)},d_{(5)},d_{(7)}\}$};

\node[font=\footnotesize] at (6.6,-1.25)
{$S_2=\{d_{(2)},d_{(5)},d_{(9)}\}$};

\node[view] (view1) at (2.4,-2.0) {
\textbf{cache view for $S_1$}\\[-0.5mm]
$\{c_q,c_{(2)},c_{(5)},c_{(7)}\}$
};

\node[view] (view2) at (6.6,-2.0) {
\textbf{cache view for $S_2$}\\[-0.5mm]
$\{c_q,c_{(2)},c_{(5)},c_{(9)}\}$
};

\draw[arrow] ($(bank.south west)!0.32!(bank.south east)$) -- (view1.north);
\draw[arrow] ($(bank.south west)!0.68!(bank.south east)$) -- (view2.north);

\node[font=\scriptsize, align=center] at (4.5,-2.70)
{shared entries: $c_q, c_{(2)}, c_{(5)}$};

\node[score] (scorer) at (4.5,-4.0) {
\textbf{same lightweight scoring rule for any candidate set}\\[1mm]
$\displaystyle
\alpha_i(S)=\frac{\exp(e_i)}{\sum_{j\in S\cup\{q\}}\exp(e_j)}
$
\qquad
$\displaystyle
s_\theta(q,S)=
w^\top \sum_{i\in S\cup\{q\}}\alpha_i(S)\,v_i
$
};

\draw[arrow] (view1.south) -- ($(scorer.north west)!0.30!(scorer.north east)$);
\draw[arrow] (view2.south) -- ($(scorer.north west)!0.70!(scorer.north east)$);


\end{tikzpicture}%
}

\caption{
\textbf{How is the cache reused?}
Each candidate set gathers the needed entries from the same cache and
applies the same lightweight scoring rule.
}
\label{fig:paraset_cache_b}
\end{subfigure}

\caption{
\textbf{Cache-based computation reuse in \ParaSet{}.}
\textbf{(a)} Expensive per-element projections and pre-softmax
compatibility terms are computed once for the query and its top-$k_b$
candidate passages.
\textbf{(b)} Different candidate sets gather the required
$c_i=(e_i,v_i)$ entries from this shared cache.
Overlapping sets therefore reuse identical cached quantities, while only
the lightweight normalization and weighted aggregation remain
subset-specific.
Thus, although the candidate space contains up to $2^{k_b}-1$ subsets,
the expensive per-element transformations are computed only once for the
$k_b$ candidate passages.
}
\label{fig:paraset_cache}
\end{figure*}

The combinatorial number of candidate sets creates a natural
computation--expressivity trade-off.
A deep cross-encoder that operates directly on the token-level query and
passage inputs can model fine-grained interactions within a set, but
repeatedly applying it during broad candidate-set exploration is expensive.
We therefore instantiate the same set-level objective with two
complementary scorers.

\subsubsection{\ParaSet{}: Efficient Set Scoring}
\label{subsection:paraset}

The passages in $\mathcal{D}_q$ have already been encoded by the
first-stage bi-encoder.
Rather than repeatedly re-encoding passage text for every candidate set,
\ParaSet{}~(Parallel Set Scorer) reuses these representations to cheaply score many different
combinations of the same passages.
Let $h_q$ denote the bi-encoder representation of the query and
$h_i$ the representation of passage $d_i$.
For a candidate set
$
S=\{d_{i_1},\ldots,d_{i_m}\},
$
\ParaSet{} forms
$
[\texttt{CLS},h_q,h_{i_1},\ldots,h_{i_m}]
$
and applies a \textit{single} self-attention layer.
The resulting \texttt{[CLS]} representation is then mapped to the
set score $s_\theta(q,S)$.

We design \ParaSet{} with a single attention layer so that repeated
scoring of overlapping candidate sets can reuse most of the computation.
Because only the resulting \texttt{[CLS]} representation is used for
scoring, attention queries are required only for the \texttt{[CLS]} token;
the query and passage elements contribute through their key and value
projections.
These projections, together with their pre-softmax interactions with the
\texttt{[CLS]} query, are independent of which other passages appear in
the candidate set and can therefore be computed once and cached.
Figure~\ref{fig:paraset_cache} illustrates this reuse.
For each element, \ParaSet{} stores the reusable quantities
$c_i=(e_i,v_i)$, where $e_i$ is its pre-softmax interaction with the
\texttt{[CLS]} query and $v_i$ is its value projection.
A candidate set then gathers only the cached entries corresponding to its
elements.
Consequently, overlapping candidate sets reuse identical per-element
computations in DP-style; only the subset-specific softmax normalization, weighted
aggregation, and scalar scoring are recomputed.
This makes broad exploration of the combinatorial set space substantially
cheaper than repeatedly applying an expressive cross-encoder to every
candidate set.

Because $S$ is modeled as an evidence \emph{set} rather than an ordered
retrieval chain, \ParaSet{} omits positional encoding for the passage
elements, making the resulting score invariant to their presentation order.
Together, the shallow architecture, cache-based reuse, and
permutation-invariant representation allow \ParaSet{} to score many
candidate sets efficiently while still modeling interactions among the
query and selected passages.
Additional implementation details are provided in
Appendix~\ref{appendix:paraset_architecture}.

\subsubsection{\texttt{SetCE}: Expressive Set Scoring}
\label{subsection:setce}

While \ParaSet{} is designed to score many candidate sets efficiently in parallel, its
shallow embedding-level interaction limits the fine-grained information
available to the scorer.
We therefore propose \texttt{SetCE}~(Set-level Cross Encoder), an
expressive cross-encoder, which jointly encodes the tokens from the query and all passages in a candidate
set, allowing interactions across the complete input. While the model architecture is different from \ParaSet{}, the same set-level compatibility learning objective is applied.
We initialize \texttt{SetCE} from a sequential cross-encoder,
\texttt{ListCE}, trained following the next-passage scoring function and
pointwise supervision of \citet{e2e}.
We then fine-tune the model using the set-level compatibility objective
from Section~\ref{subsection:set-level-compatibility-learning}.
Thus, the architecture is retained while the additional supervision is
applied directly to complete candidate sets rather than to individual
next-passage decisions.

For controlled evaluation, standalone \texttt{ListCE} and
\texttt{SetCE} use the same inference and output-selection procedure.
Their comparison therefore isolates the effect of adapting a sequentially
supervised cross-encoder with set-level compatibility learning.
Overall, \ParaSet{} and \texttt{SetCE} optimize the same notion of set
quality but serve different computational roles:
\ParaSet{} provides broad exploration, whereas
\texttt{SetCE} provides more expressive scoring for a much smaller
candidate pool.

\subsection{Efficient Inference over Candidate Sets}
\label{subsection:set-level-inference}

Even with a lightweight scorer, exhaustively enumerating every subset of
$\mathcal{D}_q$ is infeasible, where $\mathcal{D}_q$ denotes the top-$k_b$
candidate passages retrieved by the first-stage bi-encoder.
The search space contains $2^{k_b}-1$ non-empty subsets; even
$k_b=30$ already yields more than one billion candidate sets, while
$k_b=40$ yields over one trillion.
We therefore use beam search as an approximate search procedure for
maximizing $s_\theta(q,S)$ over this combinatorial set space.

\paragraph{Set-level beam-search.}
We approximately solve
$
S^\star
=
\arg\max_{\emptyset\neq S\subseteq\mathcal{D}_q}
s_\theta(q,S)
$
with a depth-limited beam search.
Let $B$ denote the beam width and $T_{\max}$ the maximum set size.

We first score all singleton sets
$\{\{d_i\}:d_i\in\mathcal{D}_q\}$ 
and retain the top-$B$ as the initial
beam.
At each subsequent depth, every retained set is expanded by adding one
passage not already contained in the set.
Importantly, the newly added passage is not scored independently;
each expanded set is reevaluated as a whole using
$s_\theta(q,S)$.
Equivalent subsets reached through different expansion orders are
deduplicated, and newly generated sets are scored.
We keep every evaluated set in an accumulated candidate pool, while only
the top-$B$ scoring sets are retained as the beam for expansion at the
next depth.

Because the most compatible evidence set may occur at any cardinality,
we do not restrict the final choice to be the deepest beam.
Instead, all evaluated candidate sets accumulated across search depths
are ranked jointly using the same $s_\theta(q,S)$.
This allows the retriever to select a final evidence set of any explored
cardinality.
The complete algorithm is provided in
Appendix~\ref{appendix:paraset_beam_search}.

\paragraph{Relation to sequential beam retrieval.}
Our inference procedure also constructs candidates incrementally by
adding passages, as in sequential beam retrieval~\citep{e2e}.
The distinction therefore does not lie in the use of beam search itself,
but in what the underlying scoring function is trained to evaluate.
Sequential retrieval supervises the relevance of a candidate next passage
conditioned on the current retrieval prefix.
Our set-level scorers instead receive direct supervision over the quality
of the resulting evidence set, including complete, incomplete, and noisy
alternatives.
Accordingly, $s_\theta(q,S)$ directly evaluates each candidate set rather
than defining its quality through an accumulation of locally supervised
next-passage decisions.

This distinction also makes scores comparable across search depths:
a candidate encountered at an earlier depth remains a valid final
solution and can be compared directly with larger sets under the same
set-level scoring function.
Set-level beam search therefore serves as an approximate search procedure over the
learned query--set compatibility function, rather than defining the
supervision itself.
A more detailed comparison is provided in
Appendix~\ref{appendix:beam_search_comparison}.

\paragraph{Set-level retrieve-and-rerank.}
Applying \texttt{SetCE} at every beam expansion would largely negate the
computational benefit of approximate exploration.
Our full \ParaSet{}$\rightarrow$\texttt{SetCE} pipeline therefore separates search
breadth from scoring capacity.

\ParaSet{} first performs the beam-search exploration described above and
produces a ranked pool of candidate sets accumulated across
cardinalities.
We retain the top-$k_p$ proposals from this pool and pass only these sets
to \texttt{SetCE}.
\texttt{SetCE} independently scores each proposal with one forward pass,
and the highest-scoring set is returned as the final retrieval output.

Thus, the beam width and maximum depth control how broadly the
combinatorial space is explored, while $k_p$ controls how many proposals
are evaluated by the expensive cross-encoder.
This factorization allows \ParaSet{} to provide inexpensive coverage of
the set space while reserving expressive \texttt{SetCE} computation for
the most promising candidates.

\section{Experiments}
\label{section:experiments}

\subsection{Experimental Setup}
\label{subsec:experimental_setup}

\paragraph{Datasets and evaluation splits.}
We evaluate on three multi-hop QA benchmarks with annotated supporting
evidence:
HotpotQA~\citep{yang2018hotpotqa},
2WikiMultihopQA~\citep{2wikimultihopqa},
and MuSiQue~\citep{musique}.
For all three benchmarks, the publicly released test split does not expose
the full gold evidence annotations required for our local retrieval
evaluation.
We therefore use the official development split of each dataset as our
final evaluation split.

To avoid tuning on this evaluation split, we hold out 500 examples from
the original training split of each dataset as an \emph{internal
development set}.
The remaining training examples are used for model training, while all
model selection and retrieval hyperparameter tuning are performed
exclusively on the internal development set.

\paragraph{Retrieval and answer generation.}
We use Contriever~\citep{contriever} and
Qwen3-Embedding-0.6B~\citep{qwen3-embedding},
denoted as Qwen3-Emb, as two first-stage bi-encoder retrievers.
For each query, the bi-encoder retrieves an initial ranked candidate pool,
which is subsequently processed by the corresponding passage-level,
sequential, or set-level retrieval method.

For downstream evaluation, the final retrieved evidence is provided to a
frozen Qwen3-8B model for answer generation.
We use the same prompting and generation configuration for all retrieval
methods, and the answer generator is not fine-tuned on any of the
evaluated datasets.

\paragraph{Compared methods.}
\textbf{Bi} directly returns passages ranked by the first-stage
bi-encoder.
\textbf{CE} is our trained passage-level cross-encoder.
It uses the same DeBERTa-v3-base~\citep{debertav3} backbone as
\texttt{ListCE} and \texttt{SetCE}, but is trained only with passage-level
relevance supervision.
Training details are provided in
Appendix~\ref{appendix:ce_training}.
\textbf{ListCE} is our sequential baseline following the next-passage
scoring function and pointwise supervision of \citet{e2e}.
\textbf{SetCE} is initialized from the trained \texttt{ListCE} scorer and
further fine-tuned with our set-level compatibility objective, while
retaining the same backbone and inference procedure as
\texttt{ListCE}.
Thus, CE, \texttt{ListCE}, and \texttt{SetCE} use the same cross-encoder
backbone and primarily differ in their scoring unit and supervision.
Finally,
\textbf{ParaSet$\rightarrow$SetCE} is our full set-level
retrieve-and-rerank pipeline:
\ParaSet{} performs lightweight candidate-set exploration, after which
\texttt{SetCE} reranks the proposed sets.
We denote this pipeline as
\textbf{P$\rightarrow$S} for short.
Unless otherwise specified, passage-level methods return their five
highest-ranked passages to the answer generator.
In Section~\ref{subsec:combined_retrieval}, we additionally use
\textbf{CE$_{\text{Electra}}$}, an off-the-shelf  cross-encoder
\citep{msmarco-electra-crossencoder}, to construct a passage-level
model-diversity control.

\paragraph{Evaluation and hyperparameter selection.}
We use downstream QA Exact Match (EM) as our primary end-to-end metric.
For retrieval-side model selection, we use \emph{Perfect Recall}, defined
as the fraction of queries for which the final retrieved output contains
all gold evidence passages.
Among the retrieval metrics considered in our analysis, Perfect Recall
shows the strongest association with downstream QA performance;
the full analysis is provided in
Appendix~\ref{appendix:retrieval_metric_correlation}.

All retrieval hyperparameters are selected exclusively on the internal
development set using Perfect Recall.
The search spaces, selected values, and sensitivity analyses are provided
in Appendix~\ref{appendix:hyperparameter_sensitivity}.

\subsection{Comparison of Retrieval Paradigms}
\label{subsec:standalone_performance}


\begin{figure*}[t]
\centering

\begin{minipage}[t]{0.6\textwidth}
\vspace{0pt}
\centering

\scriptsize
\setlength{\tabcolsep}{2.1pt}
\renewcommand{\arraystretch}{1.08}

\resizebox{\linewidth}{!}{%
\begin{tabular}{llccccc|ccc}
\toprule

&
&
\multicolumn{5}{c|}{\textbf{Standalone retrieval}}
&
\multicolumn{3}{c}{\textbf{Combined retrieval / controls}}
\\
\cmidrule(lr){3-7}
\cmidrule(lr){8-10}

\makecell{\textbf{First-stage}\\\textbf{retriever}}
&
\textbf{Dataset}
&
\textbf{Bi}
&
\textbf{CE}
&
\textbf{ListCE}
&
\textbf{SetCE}
&
\textbf{P$\rightarrow$S}
&
\textbf{CE@10}
&
\makecell{\textbf{CE} $\cup$\\\textbf{CE$_{\text{Electra}}$}}
&
\makecell{\textbf{P$\rightarrow$S}\\$\cup$ \textbf{CE}}
\\

\multicolumn{2}{r}{\textit{Avg. docs}}
&
5.00
&
5.00
&
6.39
&
7.40
&
3.49
&
10.00
&
7.71
&
7.08
\\

\midrule

\multirow{3}{*}{Contriever}
& HotpotQA
& 35.1
& 37.9
& 35.1
& 38.0
& \textbf{39.1}
& 39.5
& 40.7
& \textbf{42.2}
\\

& 2Wiki
& 20.4
& 23.8
& 17.2
& 18.4
& \textbf{23.9}
& 27.9
& 28.1
& \textbf{29.6}
\\

& MuSiQue
& 4.8
& 5.4
& 5.1
& \textbf{6.1}
& 5.8
& 5.3
& 5.9
& \textbf{6.0}
\\

\midrule

\multirow{3}{*}{Qwen3-Emb}
& HotpotQA
& 35.5
& 37.7
& 35.9
& 37.8
& \textbf{38.9}
& 39.0
& 40.6
& \textbf{41.4}
\\

& 2Wiki
& 25.0
& 26.8
& 25.2
& 26.6
& \textbf{27.4}
& 30.1
& 29.3
& \textbf{33.5}
\\

& MuSiQue
& 11.3
& 10.7
& 10.6
& \textbf{11.9}
& \textbf{11.9}
& 12.2
& 12.8
& \textbf{13.7}
\\

\bottomrule
\end{tabular}%
}

\vspace{0.3em}
\captionof{table}{
Downstream QA EM (\%) for standalone and combined retrieval.
\textit{Avg. docs} reports the average number of passages in the final
evidence context.
Bold indicates the best result within each retrieval block.
}
\label{tab:overall_results}

\end{minipage}
\hfill
\begin{minipage}[t]{0.38\textwidth}
\vspace{0pt}
\centering

\includegraphics[
    width=\linewidth
]{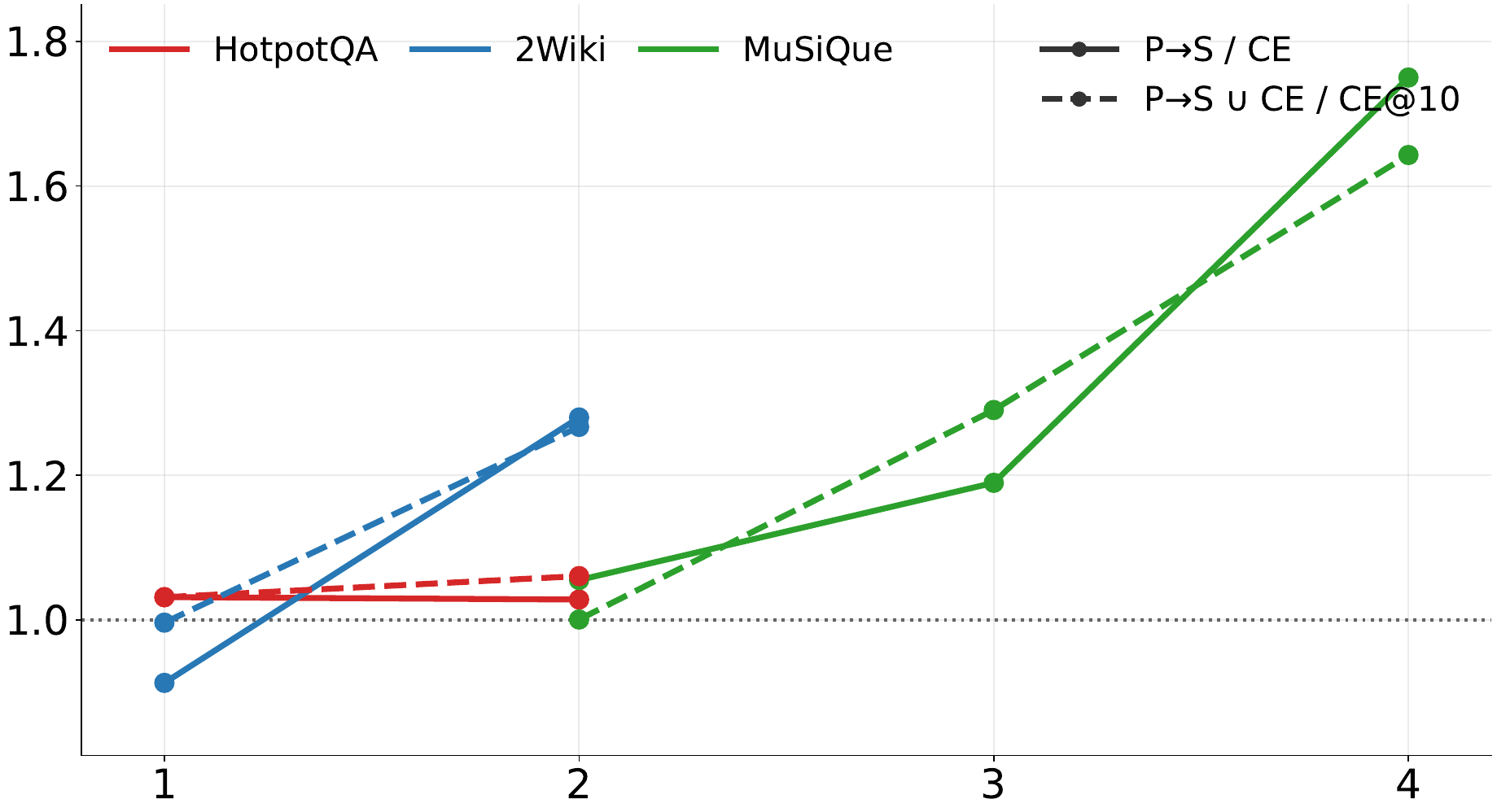}

\vspace{-0.4em}
\captionof{figure}{
Relative QA EM by EER, averaged over the two first-stage retrievers.
Solid lines denote P$\rightarrow$S / CE, and dashed lines denote
(P$\rightarrow$S $\cup$ CE) / CE@10.
}
\label{fig:eer_analysis}

\end{minipage}

\end{figure*}

\paragraph{Overall QA performance.}
We first focus on the standalone retrieval methods in the left block of
Table~\ref{tab:overall_results}.
Further fine-tuning \texttt{ListCE} with our set-level objective improves
\texttt{SetCE} in all six dataset--retriever settings.
Because the two models share the same architecture and inference procedure,
this comparison provides a controlled evaluation of the additional
set-level supervision.
This advantage is also reflected in a controlled robustness analysis,
where \texttt{SetCE} degrades more gracefully than \texttt{ListCE} under
longer and noisier retrieval prefixes
(Appendix~\ref{appendix:listce_robustness_analysis}).

Standalone \texttt{SetCE}, however, does not uniformly outperform CE,
reflecting the additional difficulty of searching the combinatorial set
space with an expressive cross-encoder.
Our full \ParaSet{}$\rightarrow$\texttt{SetCE} pipeline instead uses the
lightweight \ParaSet{} scorer for broader candidate-set exploration and
applies \texttt{SetCE} only to the resulting proposals.
The resulting pipeline outperforms CE in all six settings despite
returning only $3.49$ passages on average, compared with the fixed five
passages returned by CE.
Additional analyses of inference efficiency and proposal quality are
provided in Appendices~\ref{appendix:retrieval_efficiency}
and~\ref{appendix:paraset_proposal_quality}.

\subsection{When Does Set-Level Retrieval Help?}
\label{subsec:evidence_expansion_ratio}

Aggregate QA performance alone does not reveal whether passage-level and
set-level retrieval benefit from the same evidence structures.
We therefore characterize each example using the
\emph{Evidence Expansion Ratio} (EER),
$
\mathrm{EER}(q)
=
\frac{|\mathcal{G}_q|}
{|\mathcal{G}^{\mathrm{direct}}_q|},
$
where $\mathcal{G}_q$ is the gold evidence set and
$\mathcal{G}^{\mathrm{direct}}_q$ contains the required passages directly
connected to the query.
An EER of $1$ indicates that all required evidence is directly connected
to the query, while larger values indicate that a smaller fraction is
directly connected.
The examples of EER are illustrated in Figure~\ref{fig:eer_illustration}, and dataset-specific construction is described in
Appendix~\ref{appendix:evidence_expansion_ratio_details}.

The solid curves in Figure~\ref{fig:eer_analysis} report the relative QA EM
of P$\rightarrow$S over CE across EER levels, macro-averaged over the two
first-stage retrievers.
The relative benefit of set-level retrieval tends to increase with EER:
as a smaller fraction of the required evidence is directly connected to
the query, set-level compatibility becomes relatively more useful than
passage-level query--passage relevance.
This pattern suggests that passage-level and set-level retrieval capture
complementary aspects of multi-hop evidence structure.

\subsection{Combining Passage-Level and Set-Level Retrieval}
\label{subsec:combined_retrieval}

Motivated by this structural complementarity, we combine the output of
\ParaSet{}$\rightarrow$\texttt{SetCE} with the output of CE,
followed by deduplication.
We compare against two passage-level controls.
\textbf{CE@10} returns the ten highest-scoring passages from the same CE,
testing whether improvements arise simply from increased retrieval depth.
\textbf{CE~$\cup$~CE$_{\text{Electra}}$} combines the top-five outputs of
two independently parameterized passage-level cross-encoders, testing
whether the gain can instead be explained by passage-level model diversity.

We now turn to the right block of Table~\ref{tab:overall_results}.
The combined passage--set system outperforms both passage-level controls in
all six dataset--retriever settings.
The gain is not explained simply by returning more passages:
(P$\rightarrow$S)~$\cup$~CE contains only $7.08$ unique passages on
average, compared with ten for CE@10.
Its consistent improvement over
CE~$\cup$~CE$_{\text{Electra}}$ further indicates that the gain cannot be
explained solely by combining multiple passage-level relevance models.

The dashed curves in Figure~\ref{fig:eer_analysis} provide a complementary
view of this result.
The relative gain of the combined passage--set system over CE@10 also
tends to increase with EER, broadly following the trend observed in the
standalone comparison.
This pattern is consistent with the interpretation that combining
passage-level relevance and set-level compatibility is particularly useful
when a smaller fraction of the required evidence is directly connected to
the query.
\subsection{Comparison with LLM-Assisted Retrieval}
\label{subsec:llm_assisted_retrieval}

We finally compare against retrieval pipelines that explicitly invoke an
LLM during evidence selection.
For a controlled first-stage comparison, all methods in this section use Contriever.
We consider \textbf{IRCoT}~\citep{ircot}, which alternates LLM reasoning
and retrieval, and a \textbf{SetR-CoT\&IRI-style} selector based on the
CoT and information-requirement-identification strategy of
SetR~\citep{setr}.
The SetR-style baseline is implemented as a prompt-only selector without
task-specific fine-tuning.
Detailed implementation settings for both baselines are provided in
Appendix~\ref{appendix:llm_retrieval}.

\begin{table*}[t]
\centering
\scriptsize
\setlength{\tabcolsep}{3.2pt}
\renewcommand{\arraystretch}{1.10}

\begin{tabular}{lccccccc}
\toprule
&
\multicolumn{4}{c}{\textbf{QA EM (\%)}}
&
\multicolumn{3}{c}{\textbf{Computational Cost}}
\\
\cmidrule(lr){2-5}
\cmidrule(lr){6-8}

\multirow{2}{*}{\textbf{Method}}
&
\multirow{2}{*}{\textbf{HotpotQA}}
&
\multirow{2}{*}{\textbf{2Wiki}}
&
\multirow{2}{*}{\textbf{MuSiQue}}
&
\multirow{2}{*}{\makecell{\textbf{Macro} \textbf{Avg.}}}
&
\multicolumn{2}{c}{\textbf{Evidence Selection}}
&
\multirow{2}{*}{\makecell{\textbf{End-to-End}\\\textbf{Time (s/q)}}}
\\
\cmidrule(lr){6-7}

&
&
&
&
&
\textbf{Time (s/q)}
&
\makecell{\textbf{Peak Mem} \textbf{(GB)}}
&
\\
\midrule

IRCoT
& 36.29
& \textbf{34.89}
& \underline{5.92}
& 25.70
& 6.18
& 22.23
& 8.32
\\

SetR-CoT\&IRI-style
& \textbf{43.8}
& 26.6
& 5.8
& 25.40
& 2.10
& 21.86
& 4.25
\\

\makecell[l]{(P$\rightarrow$S) $\cup$ CE}
& \underline{42.2}
& \underline{29.6}
& \textbf{6.0}
& \textbf{25.93}
& \textbf{0.421}
& \textbf{4.96}
& \textbf{2.641}
\\

\bottomrule
\end{tabular}

\caption{
Comparison with LLM-assisted evidence selection.
Evidence-selection time is measured from receiving the query until the
final evidence context is ready for the shared answer generator, and
evidence-selection peak memory is the maximum GPU memory used during this
stage.
End-to-end time additionally includes answer generation with the shared
Qwen3-8B model.
}
\label{tab:llm_assisted_comparison}
\end{table*}

Table~\ref{tab:llm_assisted_comparison} shows that the relative strengths
of the LLM-assisted methods vary across datasets:
SetR-CoT\&IRI-style performs best on HotpotQA, while IRCoT performs best
on 2WikiMultihopQA.
Our learned retrieval pipeline ranks within the top two on all three
benchmarks and achieves the highest macro-average EM, despite requiring no
LLM during retrieval.
It also requires only $0.421$ seconds per query and $4.96$ GB of peak GPU
memory during evidence selection, corresponding to approximately
$5.0\times$ lower latency and $4.4\times$ lower memory than
SetR-CoT\&IRI-style.
Including the shared answer-generation stage, our pipeline remains
$1.6\times$ faster than SetR-CoT\&IRI-style
and $3.2\times$ faster than IRCoT
.
These results demonstrate a favorable accuracy--efficiency trade-off for
learned set-level retrieval without retrieval-time LLM inference.

\section{Conclusion}
\label{section:conclusion}

We studied multi-hop retrieval from a set-level perspective, where the
retrieval objective is to evaluate the compatibility between a query and
an evidence set rather than scoring passages independently or decomposing
retrieval into locally supervised decisions.
We introduced a set-level compatibility objective and instantiated it with two complementary models:
\ParaSet{}, an efficient candidate-set exploration,
and \texttt{SetCE}, an expressive set-level reranking.
Their combination provides a practical retrieve-and-rerank framework for
searching over a combinatorial evidence space.

Our experiments further show that set-level compatibility and passage-level
relevance capture complementary aspects of multi-hop evidence.
The relative benefit of set-level retrieval becomes larger when fewer
required passages are directly connected to the query, and combining
set-level retrieval with a passage-level cross-encoder improves downstream
QA beyond both deeper passage-level retrieval and an ensemble of different passage-level retrievers.

\newpage
\subsection*{AI use statement}

In this work, we used generative AI tools to refine experimental
analyses, and assist with the interpretation of empirical results.
We did not use generative AI tools to generate synthetic datasets,
formulate or prove mathematical claims, 
clean or reformat datasets, or conduct qualitative or thematic data
analysis; theoretical-model and proof-related uses are not applicable to
this work.

Additionally, we used generative AI tools drafting and revising portions of the
manuscript, improving readability,
and assisting with LaTeX, table, and figure formatting.
All AI-assisted suggestions, analyses, references, and manuscript text
were reviewed by the authors.
Experimental results reported in the paper were obtained from our own
experimental pipeline and were checked against the underlying outputs.
The authors made all final decisions regarding the research methodology,
experimental design, interpretation, and presentation, and take
responsibility for the complete content of the paper.





\bibliography{iclr2027_conference}
\bibliographystyle{iclr2027_conference}

\appendix

\section{Details of Set-Level Compatibility Learning}
\label{appendix:set_candidate_construction}

\subsection{Candidate-Set Construction}

For a query $q$, let
\[
\mathcal{G}_q=\{d_1^+,\ldots,d_m^+\}
\]
denote its gold evidence set.
We construct training candidates by applying three types of structured perturbation.

\paragraph{Addition.}
We introduce an irrelevant passage while retaining all required evidence:
\[
S_{\mathrm{add}}
=
\mathcal{G}_q\cup\{d^-\},
\qquad
d^-\notin\mathcal{G}_q.
\]
These examples preserve complete evidence coverage but introduce unnecessary context.

\paragraph{Elimination.}
We remove one or more required passages:
\[
\emptyset
\neq
S_{\mathrm{elim}}
\subset
\mathcal{G}_q.
\]
These examples model incomplete evidence sets.

\paragraph{Interchange.}
We replace required evidence with a distractor:
\[
S_{\mathrm{int}}
=
(\mathcal{G}_q\setminus A)\cup B,
\]
where
$A\subseteq\mathcal{G}_q$ is non-empty and
$B$ contains passages outside $\mathcal{G}_q$.
These examples jointly reduce evidence coverage and introduce noise.

We additionally use in-batch negatives constructed from gold passages associated with other queries and from subsets sampled from passages appearing in the mini-batch.

Candidate construction does not require enumerating every possible perturbation.
For each query, we sample a bounded number of candidates from each category, so the number of training examples remains controlled even as the size of the gold evidence set increases.

\subsection{Example of the Set-Quality Ordering}
\label{appendix:set_ordering_example}

Suppose
\[
\mathcal{G}_q=\{d_1^+,d_2^+\}.
\]
Consider the following candidate sets:
\[
\begin{aligned}
S_1 &= \{d_1^+,d_2^+\},\\
S_2 &= \{d_1^+,d_2^+,d^-\},\\
S_3 &= \{d_1^+\},\\
S_4 &= \{d_1^+,d^-\},\\
S_5 &= \{d^-\}.
\end{aligned}
\]

Their recall--precision pairs are
\[
\begin{aligned}
S_1 &: (1,1),\\
S_2 &: (1,2/3),\\
S_3 &: (1/2,1),\\
S_4 &: (1/2,1/2),\\
S_5 &: (0,0).
\end{aligned}
\]

Under our recall-first lexicographic ordering,
\[
S_1 \succ S_2 \succ S_3 \succ S_4 \succ S_5.
\]

The first term of our training objective requires $S_1$ to outrank every other candidate.
The fine-grained term additionally encourages neighboring quality levels to be separated, e.g.,
$S_2 \succ S_3$ and $S_3 \succ S_4$.

\section{Additional Implementation Details of \ParaSet{}}
\label{appendix:paraset_architecture}

Given a query embedding $q\in\mathbb{R}^d$ and passage embeddings
$\{d_i\}_{i\in S}$, we construct
\[
x(S)
=
[\texttt{CLS}, q, \{h_i\}_{i\in S}].
\]
A single multi-head self-attention layer followed by a lightweight
feed-forward block produces $h_{\texttt{CLS}}(S)$, and the final set score is
\[
s_\theta(q,S)
=
w^\top h_{\texttt{CLS}}(S).
\]

For each attention head, the query, key, and value projections are
\[
Q_j=W_Qx_j,
\qquad
K_j=W_Kx_j,
\qquad
V_j=W_Vx_j.
\]
Since only the \texttt{[CLS]} output is used for scoring, only its query
projection is required.
For the remaining elements, \ParaSet{} caches the key/value projections
and their pre-softmax interaction with the \texttt{[CLS]} query, as
illustrated in Figure~\ref{fig:paraset_cache}.
For each candidate subset, only the corresponding cached entries are
gathered and the subset-specific normalization and aggregation are
recomputed.

\section{Beam Search for Candidate-Set Exploration}
\label{appendix:paraset_beam_search}

Algorithm~\ref{alg:paraset_beam} describes the beam-search procedure used
by \ParaSet{} to approximate search over the combinatorial candidate-set
space.
The search maintains a beam of high-scoring sets at each cardinality while
accumulating every candidate set evaluated during the search.
The accumulated candidates are finally ranked by the same query--set
score, allowing the output to contain sets of different sizes.
When \ParaSet{} is used in the full
\ParaSet{}$\rightarrow$\texttt{SetCE} pipeline, the top-$k_p$ candidates from this
ranking are passed to \texttt{SetCE} for reranking. We used beam width $B=2$, maximum set size $T_{\max}=20$.

\begin{algorithm}[t]
\caption{Beam search for set-level candidate exploration}
\label{alg:paraset_beam}
\begin{algorithmic}[1]
\REQUIRE Query $q$, candidate pool $\mathcal{D}_q$,
beam width $B$, maximum set size $T_{\max}$

\STATE $\mathcal{S}_{\mathrm{acc}}\gets\emptyset$

\STATE $\mathcal{C}\gets\{\{d\}:d\in\mathcal{D}_q\}$
\STATE Score all $S\in\mathcal{C}$ in parallel
\STATE $\mathcal{S}_{\mathrm{acc}}
\gets
\mathcal{S}_{\mathrm{acc}}\cup\mathcal{C}$
\STATE $\mathcal{B}\gets\operatorname{TopB}(\mathcal{C})$

\FOR{$t=2,\ldots,T_{\max}$}
    \STATE $\mathcal{C}\gets\emptyset$
    \FOR{each $S\in\mathcal{B}$}
        \FOR{each $d\in\mathcal{D}_q\setminus S$}
            \STATE $\mathcal{C}\gets\mathcal{C}\cup\{S\cup\{d\}\}$
        \ENDFOR
    \ENDFOR
    \STATE Deduplicate $\mathcal{C}$
    \STATE Score all $S\in\mathcal{C}$ in parallel
    \STATE $\mathcal{S}_{\mathrm{acc}}
    \gets
    \mathcal{S}_{\mathrm{acc}}\cup\mathcal{C}$
    \STATE $\mathcal{B}\gets\operatorname{TopB}(\mathcal{C})$
\ENDFOR

\STATE \textbf{return}
$\operatorname{Rank}(\mathcal{S}_{\mathrm{acc}})$
\end{algorithmic}
\end{algorithm}

\section{Relation to Sequential Beam Retrieval}
\label{appendix:beam_search_comparison}

\ParaSet{}, \texttt{SetCE} and sequential multi-hop retrievers can all use beam search to avoid exhaustive enumeration.
However, the role of beam search and the learned scoring functions are different.

\begin{table}[t]
\centering
\small
\begin{tabular}{p{0.25\linewidth}p{0.33\linewidth}p{0.33\linewidth}}
\toprule
&
\textbf{Sequential ListCE}
&
\textbf{\ParaSet{}}\\
\midrule
Search state
&
Retrieval prefix / chain
&
Candidate passage subset
\\

Training signal
&
Pointwise next-passage supervision
&
Set-quality ranking
\\

Scoring semantics
&
Whether a candidate should extend the current prefix
&
Compatibility of the complete candidate set $s_\theta(q,S)$
\\

Training examples
&
Prefix--candidate pairs
&
Complete and perturbed passage sets
\\

Role of beam search
&
Approximate sequential retrieval
&
Approximate search over the combinatorial subset space
\\

Equivalent expansion orders
&
Represent different retrieval trajectories
&
Deduplicated as the same candidate subset
\\
\bottomrule
\end{tabular}
\caption{
Conceptual comparison between sequential Beam Retrieval-style scoring and our set-level search.
Both employ beam search as a computational primitive, but learn and optimize different scoring functions.
}
\label{tab:beam_search_comparison}
\end{table}

Our \texttt{ListCE} baseline follows the sequential cross-encoder scorer and pointwise next-passage supervision of \citet{e2e}.
Because the datasets considered in our experiments contain variable-size evidence sets, we adapt the final hypothesis selection to our common evaluation setting rather than assuming a known evidence length.
We apply the same output-selection procedure to standalone \texttt{ListCE} and \texttt{SetCE}.
Consequently, their comparison controls for architecture and inference and isolates the effect of replacing local next-passage supervision with set-level compatibility learning.

In contrast, \ParaSet{} is not trained from beam-search trajectories.
Its scorer is trained directly on complete and perturbed candidate sets.
Beam search is used only at inference time to approximate optimization of
\[
\arg\max_S s_\theta(q,S).
\]

\section{Robustness to Variable-Length and Noisy Prefixes}
\label{appendix:listce_robustness_analysis}

\begin{figure}[t]
    \centering
    \includegraphics[width=\columnwidth]{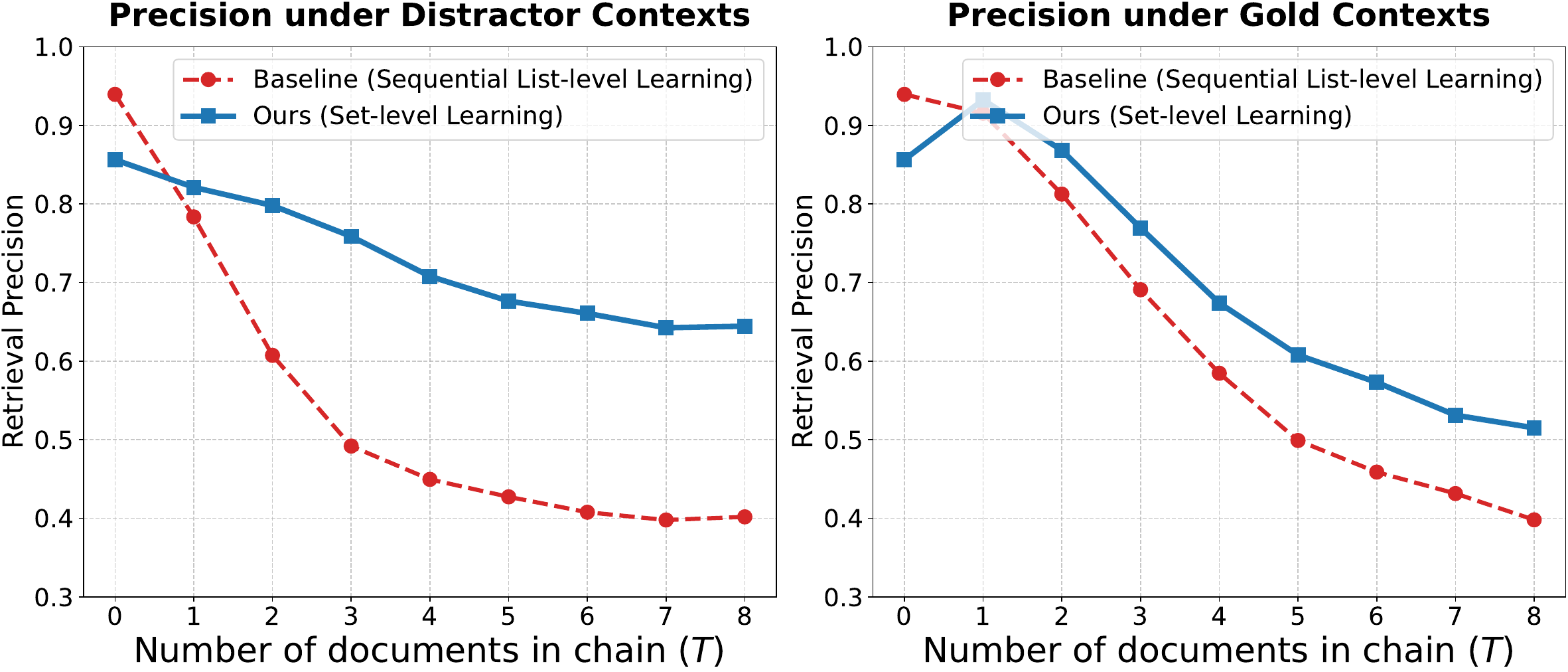}
    \caption{
    Controlled robustness analysis on MuSiQue comparing
    \texttt{ListCE} and \texttt{SetCE}, which share the same architecture
    and inference procedure but differ in supervision.
    Left: retrieval prefixes consist only of distractor passages.
    Right: prefixes contain partial gold evidence and, for sufficiently
    long prefixes, additional distractors.
    Set-level compatibility learning yields more stable retrieval precision
    as the prefix becomes longer and noisier.
    }
    \label{fig:beamce_robustness_limitation}
\end{figure}

We further examine how set-level compatibility learning affects robustness
to variable-length and partially corrupted retrieval prefixes.
We compare \texttt{ListCE}, trained with locally supervised
next-passage scoring following \citet{e2e}, with \texttt{SetCE}, obtained
by further fine-tuning the same model with our set-level compatibility
objective.
The two models therefore use the same architecture and inference procedure,
isolating the effect of set-level supervision.

Figure~\ref{fig:beamce_robustness_limitation} reports results on MuSiQue
under two controlled prefix settings.
In the \emph{distractor-only} setting, the prefix consists entirely of
irrelevant passages.
In the \emph{partially-gold} setting, for an $H$-hop question, the prefix
contains gold evidence while $T < H$; when $T \ge H$, we retain
$H-1$ gold passages and fill the remaining positions with distractors,
leaving an unseen gold passage as the retrieval target.
We then measure the model's ability to retrieve the target gold passage
as the prefix length $T$ increases.

\texttt{ListCE} degrades substantially as prefixes become longer,
particularly when distractors are introduced.
In contrast, \texttt{SetCE} remains more stable across both controlled
settings.
These results complement the end-to-end comparison in
Table~\ref{tab:overall_results}, suggesting that direct set-level
supervision improves robustness when the current evidence context varies
in length or contains irrelevant passages.

\section{Retrieval Efficiency}
\label{appendix:retrieval_efficiency}

\begin{figure*}[t]
    \centering
    \includegraphics[width=0.95\textwidth]{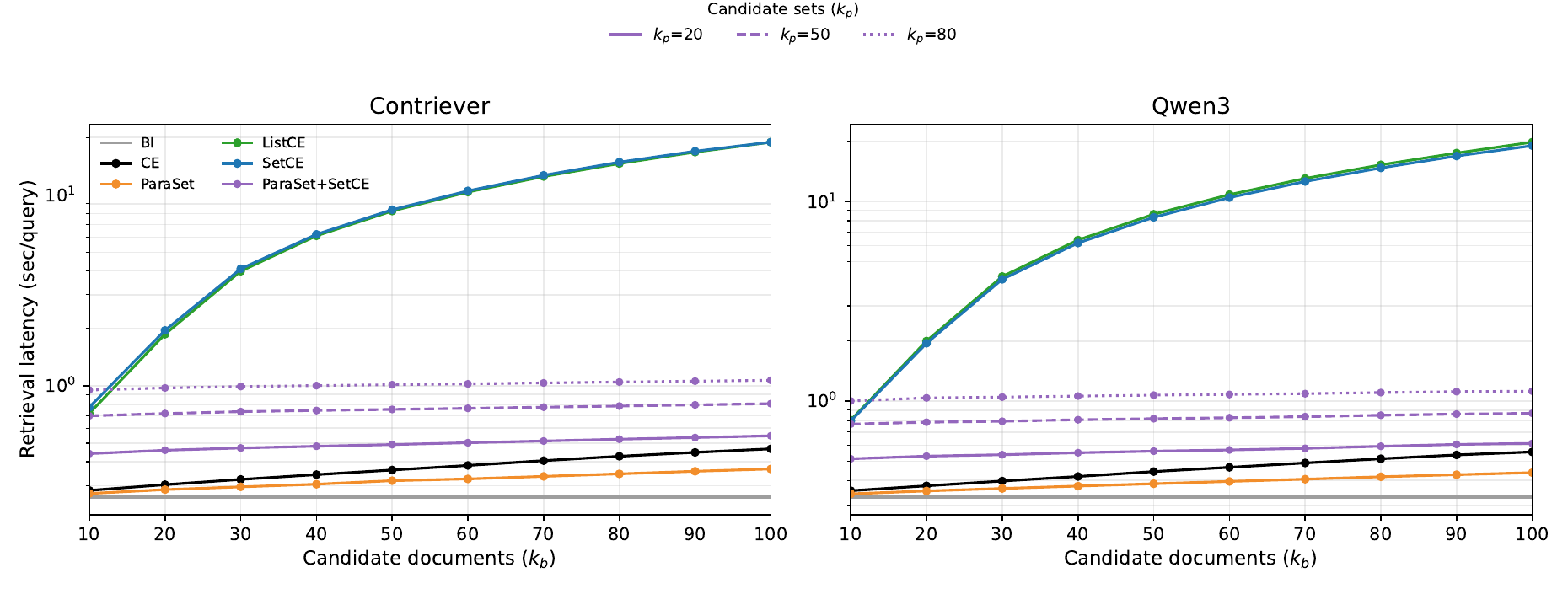}
    \caption{
    Retrieval latency as a function of the first-stage candidate-pool
    size $k_b$ for Contriever and Qwen3 in 2WikiMultihopQA dataset.
    The $y$-axis is shown on a logarithmic scale.
    Standalone \texttt{ListCE} and \texttt{SetCE} repeatedly apply a
    cross-encoder during beam expansion, while
    \ParaSet{}$\rightarrow$\texttt{SetCE} uses \ParaSet{} for candidate-set
    exploration and applies \texttt{SetCE} only to the proposed
    top-$k_p$ sets.
    }
    \label{fig:retrieval_latency}
\end{figure*}

Figure~\ref{fig:retrieval_latency} examines how retrieval latency scales
with the size of the first-stage candidate pool $k_b$.
The comparison between standalone \texttt{SetCE} and
\ParaSet{}$\rightarrow$\texttt{SetCE} isolates the computational role of the
retrieve-and-rerank factorization.
Standalone \texttt{SetCE} uses the cross-encoder throughout iterative
beam expansion, whereas the full pipeline performs combinatorial
exploration with the lightweight \ParaSet{} scorer and invokes
\texttt{SetCE} only after exploration is complete.

The two inference strategies exhibit markedly different scaling behavior.
For both first-stage retrievers, the latency of standalone \texttt{SetCE}
increases steeply as $k_b$ grows, from $0.6$ sec/query at
$k_b=10$ to $20$ sec/query at $k_b=100$.
\texttt{ListCE} exhibits nearly identical latency, as expected from its
shared architecture and inference procedure with \texttt{SetCE}.
In contrast, \ParaSet{}$\rightarrow$\texttt{SetCE} scales much more slowly and
remains below $0.4$ sec/query even at $k_b=100$.
At this candidate-pool size, the retrieve-and-rerank pipeline is therefore
more than $50\times$ faster than standalone \texttt{SetCE} under both
retrieval backbones.

This difference highlights the computational role of \ParaSet{}.
The main bottleneck of standalone \texttt{SetCE} is not the final
evaluation of a candidate set itself, but repeatedly applying an
expressive text cross-encoder while exploring a rapidly growing
combinatorial search space.
\ParaSet{} moves this exploration to a shallow scorer operating on cached
bi-encoder representations, after which the expensive \texttt{SetCE}
scorer is applied only to a small selected pool.
Thus, the two models play the same complementary roles as retrieval and
reranking in conventional passage-level pipelines, but over candidate
evidence sets.

\section{Training Details of the Passage-Level Cross-Encoder}
\label{appendix:ce_training}

Our primary passage-level baseline, CE, uses the same
DeBERTa-v3-base backbone as \texttt{ListCE} and \texttt{SetCE},
but is trained only to score the relevance of individual
query--passage pairs.

\paragraph{Training instances.}
For each query, we construct training instances for each passage in its gold evidence set, treating each gold passage in turn as the positive passage.
We pair this positive with 15 negative passages.
By default, approximately half of the negatives are sampled from
non-gold passages in the query's first-stage candidate pool, while the
remaining negatives are in-batch passages associated with other queries.
The query-specific negatives are therefore retrieval-derived distractors,
but are sampled from the non-gold candidate pool rather than being
restricted to the highest-ranked candidates.

\paragraph{Passage-level ranking objective.}
Let $p^+$ denote the positive passage and
$\{p^-_1,\ldots,p^-_N\}$ the negative passages.
The cross-encoder independently produces a relevance score
$s_{\mathrm{CE}}(q,p)$ for each query--passage pair.
We minimize the softmax cross-entropy objective

\[
\mathcal{L}_{\mathrm{CE}}
=
-\log
\frac{
    \exp s_{\mathrm{CE}}(q,p^+)
}{
    \exp s_{\mathrm{CE}}(q,p^+)
    +
    \sum_{i=1}^{N}
    \exp s_{\mathrm{CE}}(q,p^-_i)
},
\]

with $N=15$ in our experiments.

This objective trains the model to rank individually relevant passages
above non-relevant alternatives.
Unlike \texttt{ListCE} and our set-level models, CE does not condition
on previously retrieved passages and receives no supervision over
passage combinations, retrieval order, or evidence-set compatibility.
At inference time, each candidate passage is scored independently and
the top-$k$ passages are returned according to
$s_{\mathrm{CE}}(q,p)$.

\section{Correlation between Retrieval Metrics and Downstream QA}
\label{appendix:retrieval_metric_correlation}

We examine which retrieval-side metric is most aligned with downstream QA performance on the internal development set.

\paragraph{Compared retrieval metrics.}
Let precision and recall of a retrieved passage set with respect to the gold evidence be denoted by $P$ and $R$, respectively.
We consider the standard $F_\beta$ family,
\[
F_\beta
=
(1+\beta^2)
\frac{PR}{\beta^2 P + R},
\]
where larger $\beta$ places greater weight on recall.
In particular, $F_1$ weights precision and recall equally, $F_0$ corresponds to precision, and the large-$\beta$ limit corresponds to recall.

We additionally consider \emph{Perfect Recall} (PR), which measures whether the complete gold evidence set is recovered:
\[
\mathrm{PR}
=
\frac{1}{|\mathcal{Q}|}
\sum_{q\in\mathcal{Q}}
\mathbbm{1}
\left[
\mathcal{G}_q \subseteq \hat{\mathcal{S}}_q
\right],
\]
where $\mathcal{G}_q$ is the gold evidence set and
$\hat{\mathcal{S}}_q$ is the retrieved output for query $q$.
Unlike ordinary recall, Perfect Recall requires \emph{all} required evidence passages to be present for an example to receive credit.

\paragraph{Correlation with downstream QA.}
For the retrieval configurations evaluated on the internal development set, we compute each retrieval metric together with downstream QA EM and measure their Spearman rank correlation.
Table~\ref{tab:retrieval_metric_correlation} reports the results.

\begin{table}[t]
\centering
\small
\setlength{\tabcolsep}{6pt}
\renewcommand{\arraystretch}{1.05}
\caption{
Spearman rank correlation between retrieval-side metrics and downstream QA Exact Match (EM) on the internal development set.
Perfect Recall exhibits the strongest correlation with QA EM.
Among the $F_\beta$ metrics, $F_5$ gives the highest correlation.
}
\label{tab:retrieval_metric_correlation}
\begin{tabular}{lc}
\toprule
\textbf{Retrieval metric}
& \textbf{Spearman $\rho$ with QA EM}
\\
\midrule
Precision ($F_0$) & 0.0912 \\
$F_1$             & 0.2596 \\
$F_2$             & 0.4506 \\
$F_3$             & 0.5526 \\
$F_4$             & 0.6538 \\
${F_5}$    & \textbf{0.7045} \\
$F_6$             & 0.6994 \\
Recall ($F_{\infty}$) & 0.6929 \\
\midrule
\textbf{Perfect Recall} & \textbf{0.7612} \\
\bottomrule
\end{tabular}
\end{table}

The correlation increases substantially as $F_\beta$ places greater weight on recall, peaking at $F_5$ among the evaluated $F_\beta$ metrics.
This confirms that downstream multi-hop QA is substantially more sensitive to evidence coverage than to precision alone.
However, Perfect Recall exhibits an even stronger correlation with QA EM ($\rho=0.7612$), exceeding both $F_5$ ($\rho=0.7045$) and ordinary recall ($\rho=0.6929$).

This result is consistent with the structure of multi-hop QA.
A retriever may achieve high passage-level recall while still missing one required piece of evidence on many individual examples.
Perfect Recall directly measures whether the complete evidence set required for answering a query has been recovered, and therefore more closely reflects the all-or-nothing coverage requirement of multi-hop reasoning.
We consequently use Perfect Recall, rather than $F_\beta$, for retrieval hyperparameter selection throughout our main experiments.

\section{Training and Hyperparameter Details}
\label{appendix:training_hyperparameters}

\subsection{Optimization and Objective Hyperparameters}
\label{appendix:optimization_details}

Table~\ref{tab:training_hyperparameters} summarizes the optimization and
set-level objective hyperparameters used for \ParaSet{} and
\texttt{SetCE}.
Both models are trained with AdamW~\citep{adamw} using the default PyTorch
$\beta$ and $\epsilon$ values, without a learning-rate scheduler.
For \ParaSet{}, we tune the adjacent-ranking weight $\lambda$ over
$\{0.1,0.2,\ldots,1.0\}$ on the internal development set. 

\begin{table}[t]
\centering
\small
\setlength{\tabcolsep}{5pt}
\renewcommand{\arraystretch}{1.08}

\begin{tabular}{lcc}
\toprule
\textbf{Hyperparameter}
& \textbf{\ParaSet{}}
& \textbf{\texttt{SetCE}}
\\
\midrule

Optimizer
& AdamW
& AdamW
\\

AdamW $(\beta_1,\beta_2)$
& $(0.9,0.999)$
& $(0.9,0.999)$
\\

AdamW $\epsilon$
& $10^{-8}$
& $10^{-8}$
\\

Learning rate
& $1\times10^{-5}$
& $5\times10^{-6}$
\\

Weight decay
& $0.01$
& $0.01$
\\

Margin $\gamma$
& $0.1$
& $0.5$
\\

Adjacent-ranking weight $\lambda$
& \makecell{$0.2$ for HotpotQA \& 2Wiki\\$0.8$ for MuSiQue}
& $0.2$
\\

\bottomrule
\end{tabular}

\caption{
Optimization and objective hyperparameters for \ParaSet{} and
\texttt{SetCE}.
For \ParaSet{}, the adjacent-ranking weight $\lambda$ is selected
separately for each dataset as described below.
}
\label{tab:training_hyperparameters}
\end{table}

\subsection{Hyperparameter Selection and Sensitivity}
\label{appendix:hyperparameter_sensitivity}

\paragraph{Selection protocol.}
All retrieval hyperparameters are selected exclusively on the internal
development set using Perfect Recall (PR), which exhibits the strongest
correlation with downstream QA performance among the retrieval-side metrics
considered in Appendix~\ref{appendix:retrieval_metric_correlation}.
For CE, we search over the bi-encoder candidate-pool size
$k_b \in \{10,20,\ldots,100\}$ and $k_b \in \{10,20,\ldots,50\}$ for ListCE and SetCE.
For \ParaSet{}$\rightarrow$\texttt{SetCE}, we additionally search over the number of
candidate sets passed to the reranker,
$k_p \in \{10,20,\ldots,100\}$,
and jointly select $(k_b,k_p)$ by development-set PR.
The selected configurations are summarized in
Table~\ref{tab:selected_hyperparameters}.
No hyperparameter is tuned on the final evaluation split.

\begin{table}[t]
\centering
\small
\setlength{\tabcolsep}{4.5pt}
\renewcommand{\arraystretch}{1.08}
\begin{tabular}{llcccc}
\toprule
\textbf{Dataset}
& \textbf{Retriever}
& \textbf{CE}
& \textbf{ListCE}
& \textbf{SetCE}
& \makecell{\textbf{P$\rightarrow$S}\\$(k_b,k_p)$}
\\
\midrule

\multirow{2}{*}{HotpotQA}
& Contriever & 80 & 10 & 30 & $(50,80)$ \\
& Qwen3      & 70 & 10 & 40 & $(30,90)$ \\

\midrule

\multirow{2}{*}{2Wiki}
& Contriever & 100 & 50 & 40 & $(90,20)$ \\
& Qwen3      & 50  & 50 & 10 & $(40,40)$ \\

\midrule

\multirow{2}{*}{MuSiQue}
& Contriever & 50 & 50 & 50 & $(20,90)$ \\
& Qwen3      & 50 & 50 & 50 & $(30,70)$ \\

\bottomrule
\end{tabular}
\caption{
Hyperparameters selected by Perfect Recall on the internal development set.
Entries for CE, \texttt{ListCE}, and \texttt{SetCE} denote the
first-stage candidate-pool size $k_b$.
For \ParaSet{}$\rightarrow$\texttt{SetCE}, we report the jointly selected
$(k_b,k_p)$, where $k_p$ is the number of candidate sets passed from
\ParaSet{} to \texttt{SetCE}.
}
\label{tab:selected_hyperparameters}
\end{table}

\begin{figure*}[t]
    \centering
    \includegraphics[width=\textwidth]
    {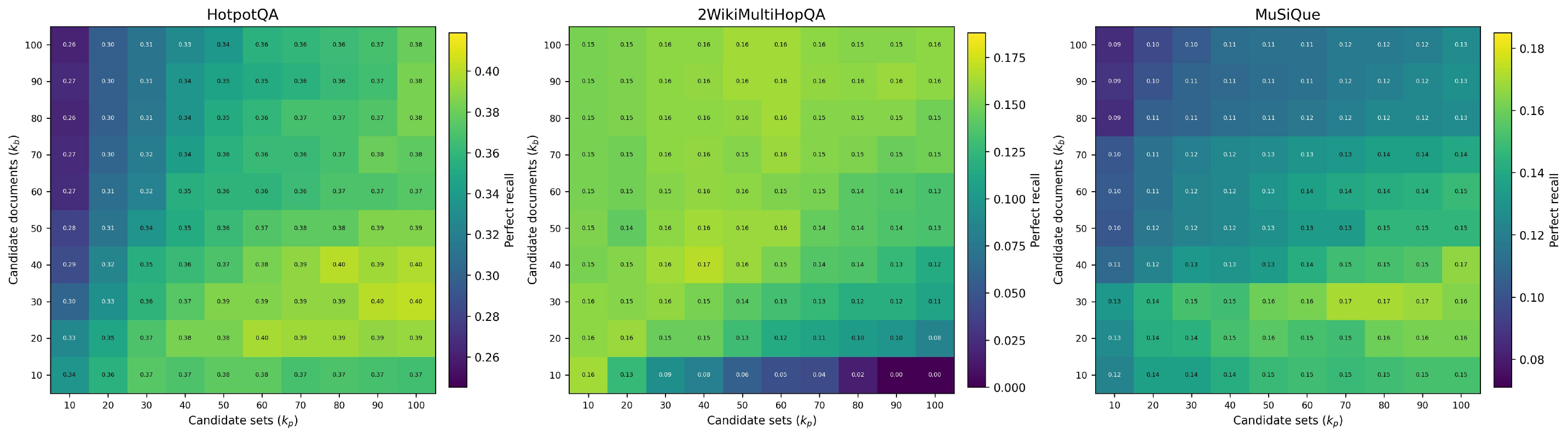}
    \caption{
    Perfect Recall of \ParaSet{}$\rightarrow$\texttt{SetCE} across the
    $(k_b,k_p)$ hyperparameter grid with Qwen3 as the first-stage retriever.
    Rows vary the number of candidate passages $k_b$, and columns vary the
    number of candidate sets $k_p$ passed from \ParaSet{} to \texttt{SetCE}.
    }
    \label{fig:paraset_setce_sensitivity}
\end{figure*}

\paragraph{Sensitivity of \ParaSet{}$\rightarrow$\texttt{SetCE}.}
Because the full retrieve-and-rerank pipeline introduces a second
hyperparameter $k_p$, we additionally examine its sensitivity jointly with
the first-stage candidate-pool size $k_b$.
Figure~\ref{fig:paraset_setce_sensitivity} reports development-set Perfect
Recall over the full $(k_b,k_p)$ grid for the Qwen3 first-stage retriever.

Across all three datasets, high-performing configurations occupy relatively
broad regions of the $(k_b,k_p)$ space rather than appearing as isolated
single-point optima.
Neighboring configurations around the selected values generally achieve
similar Perfect Recall, indicating that the performance of the pipeline is
not driven by a narrowly tuned combination of $k_b$ and $k_p$.
The location of the best region nevertheless differs across datasets,
which motivates selecting the two parameters jointly rather than fixing a
single configuration for all tasks.

The two hyperparameters also control distinct stages of the pipeline.
Increasing $k_b$ exposes the set explorer to a larger first-stage passage
pool, whereas increasing $k_p$ allows the expressive \texttt{SetCE}
reranker to examine more proposals produced by \ParaSet{}.
Consequently, larger values are not expected to yield monotonic
improvements: enlarging the passage pool can introduce additional
distractors, while increasing $k_p$ beyond a sufficiently broad proposal
set provides diminishing additional coverage.
Overall, the smooth performance surfaces indicate that the benefits of the
retrieve-and-rerank pipeline are robust to moderate variation in either
hyperparameter.

\section{Candidate-Set Proposal Quality of \ParaSet{}}
\label{appendix:paraset_proposal_quality}

The full \ParaSet{}$\rightarrow$\texttt{SetCE} pipeline relies on \ParaSet{} as a
candidate-set proposer: \ParaSet{} explores the combinatorial set space and
passes its top-$k_p$ candidate sets to the more expressive \texttt{SetCE}
reranker.
We therefore separately examine whether high-quality evidence sets are
actually retained within this proposal pool.

\paragraph{Max-PR@$k_p$.}
For a query $q$, let
$\mathcal{C}^{(k_p)}_q$ denote the top-$k_p$ candidate sets ranked by
\ParaSet{}, and let $\mathcal{G}_q$ denote the gold evidence set.
We define

\begin{equation}
\label{eq:max_pr_at_k}
\mathrm{Max\text{-}PR}@k_p
=
\frac{1}{|\mathcal{Q}|}
\sum_{q\in\mathcal{Q}}
\max_{S\in\mathcal{C}^{(k_p)}_q}
\mathbbm{1}
\left[
\mathcal{G}_q \subseteq S
\right].
\end{equation}

Thus, Max-PR@$k_p$ measures the fraction of queries for which at least one
of the top-$k_p$ sets proposed by \ParaSet{} contains the complete gold
evidence.
Equivalently, it represents the oracle Perfect Recall attainable by a
reranker whose candidate set is restricted to the top-$k_p$ proposals.
By construction, Max-PR@$k_p$ is non-decreasing in $k_p$.

\begin{figure*}[t]
    \centering
    \includegraphics[width=\textwidth]
    {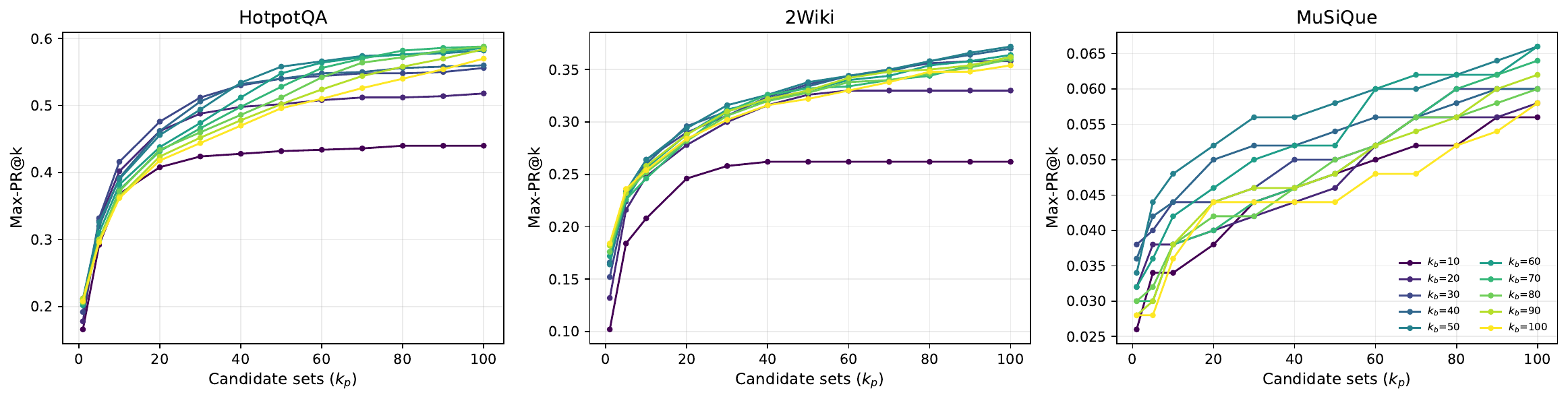}
    \caption{
    Candidate-set proposal quality of \ParaSet{} with Contriever as the
    first-stage retriever.
    We report Max-PR@$k_p$, the fraction of queries for which at least one
    of the top-$k_p$ candidate sets proposed by \ParaSet{} contains all
    gold evidence passages.
    Each curve corresponds to a different first-stage candidate-pool size
    $k_b$.
    }
    \label{fig:paraset_max_pr}
\end{figure*}

\paragraph{ParaSet recovers useful candidates within a small proposal pool.}
Figure~\ref{fig:paraset_max_pr} shows that increasing $k_p$ consistently
increases the oracle evidence coverage available to the downstream
reranker.
On HotpotQA and 2WikiMultihopQA, much of this gain is obtained within the
first few tens of proposals, after which the curves gradually saturate.
This indicates that \ParaSet{} concentrates many of the useful evidence
sets relatively high in its ranking rather than requiring exhaustive
enumeration of the candidate-set space.
MuSiQue exhibits a more gradual improvement as $k_p$ increases, suggesting
that useful complete evidence sets are more dispersed in the proposal
ranking for this more difficult retrieval setting.

\paragraph{Interaction with the first-stage candidate pool.}
The proposal quality also depends strongly on $k_b$.
When $k_b$ is small, the first-stage retriever may omit one or more required
passages, in which case no subsequent set-level search can recover a
complete evidence set.
This is particularly visible for $k_b=10$, whose curves saturate
substantially below those obtained with larger candidate pools on HotpotQA
and 2WikiMultihopQA.
Increasing $k_b$ raises the attainable proposal coverage by exposing
\ParaSet{} to a broader passage pool, although it simultaneously enlarges
the combinatorial search space and introduces additional distractors.

These results clarify the division of labor in our retrieve-and-rerank
pipeline.
\ParaSet{} does not need to identify the single best set perfectly by
itself; its role is to compress a large combinatorial space into a small
collection of promising candidate sets while preserving high-quality
evidence combinations.
\texttt{SetCE} can then devote its more expressive cross-encoder scoring
only to this reduced proposal pool.
The observed Max-PR@$k_p$ curves show that increasing $k_p$
systematically expands the evidence available to the reranker, while the
diminishing gains at larger $k_p$ motivate using a moderate, validation-selected value of $k_p$ rather than exhaustively scoring candidate sets.

\section{Dataset-Specific Construction of the Evidence Expansion Ratio}
\label{appendix:evidence_expansion_ratio_details}

We derive the Evidence Expansion Ratio (EER) solely from
dataset-provided annotations and construction metadata; retrieval outputs
and model predictions are never used.
Let $\mathcal{G}_q$ denote the gold evidence set for query $q$, and let
$\mathcal{G}^{\mathrm{direct}}_q \subseteq \mathcal{G}_q$ contain the
gold passages directly connected to the query in the annotated evidence
structure.
We define
\[
\mathrm{EER}(q)
=
\frac{
|\mathcal{G}_q|
}{
|\mathcal{G}^{\mathrm{direct}}_q|
}.
\]

Conceptually, we associate each gold evidence passage with a reasoning
step and regard a passage as \emph{directly connected} when its
corresponding step has no prerequisite gold-evidence step.
Thus, if an example contains $m$ gold passages and $r$ of them correspond
to such source steps, then
\[
\mathrm{EER}(q)=\frac{m}{r}.
\]
An EER of $1$ indicates that all required evidence is directly connected
to the query, while larger values indicate that a smaller fraction of the
required evidence is directly connected.
Because the three datasets expose their evidence structures in different
forms, we instantiate this definition using the native annotations and
construction metadata of each dataset as described below.

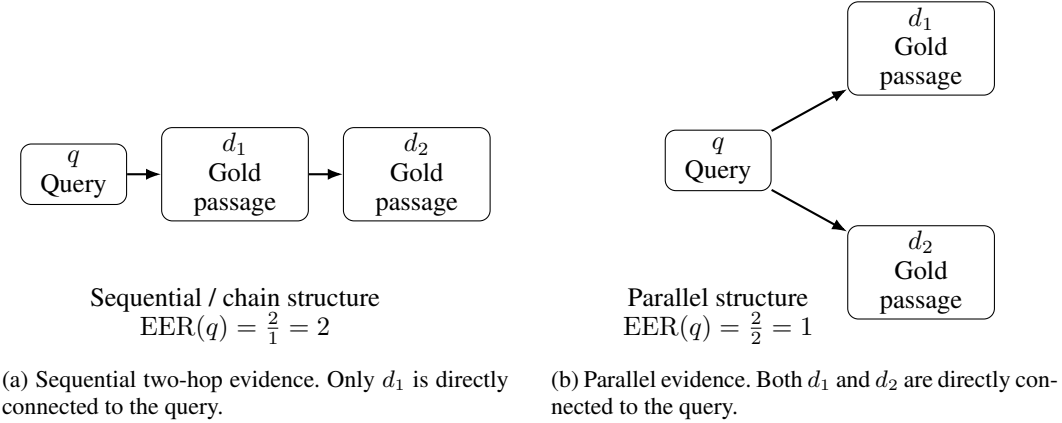
\begin{figure}[t]
    \centering

    \begin{subfigure}[t]{0.48\linewidth}
        \centering
        \begin{tikzpicture}[
            query/.style={
                draw,
                rounded corners,
                align=center,
                minimum width=1.4cm,
                minimum height=0.8cm,
                inner sep=2pt
            },
            passage/.style={
                draw,
                rounded corners,
                align=center,
                text width=1.8cm,
                minimum height=0.8cm,
                inner sep=2pt
            },
            arrow/.style={
                ->,
                thick,
                >={Latex[length=2mm]}
            }
        ]

            \node[query] (q) {$q$\\Query};
            \node[passage, right=0.45cm of q] (d1)
                {$d_1$\\Gold passage};
            \node[passage, right=0.45cm of d1] (d2)
                {$d_2$\\Gold passage};

            \draw[arrow] (q) -- (d1);
            \draw[arrow] (d1) -- (d2);

            \node[
                draw=none,
                below=0.75cm of d1,
                align=center
            ] {
                Sequential / chain structure\\
                $\mathrm{EER}(q)=\frac{2}{1}=2$
            };

        \end{tikzpicture}

        \caption{
        Sequential two-hop evidence.
        Only $d_1$ is directly connected to the query.
        }
        \label{fig:eer_sequential}
    \end{subfigure}
    \hfill
    \begin{subfigure}[t]{0.48\linewidth}
        \centering
        \begin{tikzpicture}[
            query/.style={
                draw,
                rounded corners,
                align=center,
                minimum width=1.4cm,
                minimum height=0.8cm,
                inner sep=2pt
            },
            passage/.style={
                draw,
                rounded corners,
                align=center,
                text width=1.8cm,
                minimum height=0.8cm,
                inner sep=2pt
            },
            arrow/.style={
                ->,
                thick,
                >={Latex[length=2mm]}
            }
        ]

            \node[query] (q) {$q$\\Query};

            \node[
                passage,
                above right=0.45cm and 1.0cm of q
            ] (d1)
                {$d_1$\\Gold passage};

            \node[
                passage,
                below right=0.45cm and 1.0cm of q
            ] (d2)
                {$d_2$\\Gold passage};

            \draw[arrow] (q) -- (d1);
            \draw[arrow] (q) -- (d2);

            \node[
                draw=none,
                below=1.15cm of q,
                align=center
            ] {
                Parallel structure\\
                $\mathrm{EER}(q)=\frac{2}{2}=1$
            };

        \end{tikzpicture}

        \caption{
        Parallel evidence.
        Both $d_1$ and $d_2$ are directly connected to the query.
        }
        \label{fig:eer_parallel}
    \end{subfigure}

    \caption{
    Illustration of the Evidence Expansion Ratio (EER) for two-hop
    evidence structures.
    EER is the ratio of the total number of required evidence passages
    to the number directly connected to the query.
    }
    \label{fig:eer_illustration}
\end{figure}

\paragraph{HotpotQA.}
HotpotQA provides an official question-type annotation with two
categories, \texttt{bridge} and \texttt{comparison}
~\citep{yang2018hotpotqa}, but does not provide an explicit per-example
dependency graph between the two supporting passages.
We therefore use the question type as a construction-based proxy for the
intended evidence structure.
For \texttt{comparison} questions, the two compared entities are
introduced in the original question, so we treat both supporting passages
as directly connected and assign
\[
\mathrm{EER}=\frac{2}{2}=1.
\]
For \texttt{bridge} questions, the dataset was constructed from linked
article pairs connected through a bridge entity.
We therefore treat one supporting passage as directly connected to the
query and the other as depending on the bridge established by that
passage, yielding
\[
\mathrm{EER}=\frac{2}{1}=2.
\]
This mapping reflects the intended retrieval structure of the dataset
construction rather than asserting that every \texttt{bridge} instance
necessarily follows a strict two-step reasoning chain.

\paragraph{2WikiMultiHopQA.}
2WikiMultiHopQA is generated from structured reasoning templates and
provides four question types---\texttt{compositional},
\texttt{inference}, \texttt{comparison}, and
\texttt{bridge-comparison}---together with evidence annotations
~\citep{2wikimultihopqa}.
We map these templates to their corresponding source-evidence structure.
For \texttt{compositional} and \texttt{inference} questions, one of the
two supporting passages provides intermediate information required to
reach the other.
We therefore treat one of the two passages as directly connected and
assign
\[
\mathrm{EER}=\frac{2}{1}=2.
\]
For \texttt{comparison} questions, both supporting passages correspond
to entities specified by the original question, giving
\[
\mathrm{EER}=\frac{2}{2}=1.
\]
Finally, \texttt{bridge-comparison} questions combine two bridge-style
branches before comparing their outcomes.
Among the four supporting passages, the first passage of each branch is
directly connected to the query, whereas the second depends on its
corresponding intermediate entity.
We therefore assign
\[
\mathrm{EER}=\frac{4}{2}=2.
\]

\paragraph{MuSiQue.}
MuSiQue explicitly provides the constituent single-hop decomposition from
which each multi-hop question is composed~\citep{musique}.
We construct the corresponding directed acyclic graph (DAG), with one
node for each decomposition step and its supporting passage.
We add an edge from step $j$ to step $i$ whenever step $i$ depends on the
answer produced by step $j$, represented by a reference such as
\texttt{\#j} in the decomposition.
A decomposition step with indegree zero has no prerequisite step and is
therefore treated as directly connected to the original query.
If the decomposition contains $n$ steps and $r$ source nodes, we set
\[
\mathrm{EER}(q)=\frac{n}{r}.
\]
This construction naturally accommodates chains as well as branching and
merging dependency structures without assigning a manually defined
topology label to each example.

Figure~\ref{fig:eer_illustration} illustrates the intuition behind EER
for two-hop evidence structures.
In a sequential chain, only one of the two supporting passages is directly
connected to the query, yielding $\mathrm{EER}=2$.
In contrast, in a parallel structure both supporting passages are directly
connected to the query, yielding $\mathrm{EER}=1$.

\section{Implementation Details for LLM-Assisted Retrieval Baselines}
\label{appendix:llm_retrieval}

\paragraph{IRCoT.}
We implement IRCoT~\citep{ircot} as an iterative
reasoning--retrieval pipeline.
At each retrieval step, the current reasoning context is used to query
Contriever, and the top three passages are retrieved.
Passages retrieved in previous iterations are excluded from subsequent
retrieval results, so each step contributes up to three previously unseen
passages.
We allow at most six retrieval iterations, corresponding to at most
18 unique retrieved passages.

\paragraph{SetR-CoT\&IRI-style.}
We construct an LLM-based set-selection baseline following the
CoT\&IRI prompting procedure introduced in SetR~\citep{setr}.
Specifically, we first retrieve the top-20 passages with Contriever and
provide them, together with the query, to Qwen3-8B.
We use the SetR-CoT\&IRI prompt to encourage the model to reason about
the query, identify its information requirements, and select a subset of
passages that collectively covers those requirements.
We do not perform the task-specific fine-tuning used by the original
SetR model; instead, Qwen3-8B is used off the shelf with the published
prompting procedure.
We therefore refer to this baseline as
\emph{SetR-CoT\&IRI-style} rather than a full reproduction of SetR.

\end{document}